\documentstyle[12pt,epsf]{article}

\makeatletter
\@addtoreset{equation}{section}
\makeatother

\def\BbbZ{Z}
\def\half{{\textstyle{1\over2}}}

\def\Gtt{\skew4\tilde{\tilde G}}
\def\gtt{\skew3\tilde{\tilde g}}
\def\Htt{\skew4\tilde{\tilde H}}
\def\Phitt{\skew0\tilde{\tilde\Phi}}
\def\etatt{\skew3\tilde{\tilde\eta}}
\def\Vtt{\skew2\tilde{\tilde V}}
\def\cHtt{\skew5\tilde{\tilde {\cal H}}}
\def\Ltt{\skew2\tilde{\tilde L}}
\def\Mtt{\skew4\tilde{\tilde M}}
\def\Rtt{\skew4\tilde{\tilde R}}
\def\cFtt{\skew5\tilde{\tilde {\cal F}}}
\def\Btt{\skew4\tilde{\tilde B}}
\def\Ftt{\skew4\tilde{\tilde F}}
\def\Att{\skew5\tilde{\tilde A}}
\def\Ntt{\skew4\tilde{\tilde N}}
\def\Ztt{\skew4\tilde{\tilde Z}}
\def\Ptt{\skew4\tilde{\tilde P}}
\def\Qtt{\skew4\tilde{\tilde Q}}
\def\Ett{\skew4\tilde{\tilde E}}

\let\a=\alpha \let\b=\beta \let\g=\gamma  \let\e=\epsilon

\let\s=\sigma

\let\la=\label  
  
\def\nn{\nonumber} \def\bd{\begin{document}} \def\ed{\end{document}}
\def\ds{\documentstyle} \let\fr=\frac \let\bl=\bigl \let\br=\bigr
\let\Br=\Bigr \let\Bl=\Bigl
\let\bm=\bibitem
\let\na=\nabla
\let\pa=\partial \let\ov=\overline
\newcommand{\be}{\begin{equation}}
\newcommand{\ee}{\end{equation}}
\def\ba{\begin{array}}
\def\ea{\end{array}}
\newcommand{\ho}[1]{$\, ^{#1}$}
\newcommand{\hoch}[1]{$\, ^{#1}$}
\newcommand{\bea}{\begin{eqnarray}}
\newcommand{\eea}{\end{eqnarray}}
\newcommand{\ra}{\rightarrow}
\newcommand{\lra}{\longrightarrow}
\newcommand{\Lra}{\Leftrightarrow}
\newcommand{\ap}{\alpha^\prime}
\newcommand{\bp}{\tilde \beta^\prime}
\newcommand{\tr}{{\rm tr} }
\newcommand{\Tr}{{\rm Tr} }
\newcommand{\NP}{Nucl. Phys. }
\newcommand{\tamphys}{\it
Center for Theoretical Physics, Department of Physics\\
Texas A\&M University, College Station, Texas 77843--4242}

\newcommand{\auth}{M. J. Duff, James T. Liu and J. Rahmfeld }

\begin{document}
\hfill{}

\hfill{CTP-TAMU-27/95}

\hfill{hep-th/9508094}

\vspace{24pt}

\begin{center}
{ \large {\bf FOUR DIMENSIONAL STRING/STRING/STRING TRIALITY\footnote{
Research supported in part by NSF Grant PHY-9411543.} }}

\vspace{24pt}

\auth

\vspace{10pt}

{\tamphys}

\vspace{24pt}

\underline{ABSTRACT}

\end{center}

In six spacetime dimensions, the heterotic string is dual to a Type
$IIA$ string.  On further toroidal compactification to four spacetime
dimensions, the heterotic string acquires an $SL(2,\BbbZ)_S$
strong/weak coupling duality and an $SL(2,\BbbZ)_T \times
SL(2,\BbbZ)_U$ target space duality acting on the dilaton/axion,
complex Kahler form and the complex structure fields $S,T,U$
respectively.  Strong/weak duality in $D=6$ interchanges the roles of
$S$ and $T$ in $D=4$ yielding a Type $IIA$ string with fields $T,S,U$.
This suggests the existence of a third string (whose six-dimensional
interpretation is more obscure) that interchanges the roles of $S$ and
$U$. It corresponds in fact to a Type $IIB$ string with fields $U,T,S$
leading to a four-dimensional string/string/string triality. Since
$SL(2,\BbbZ)_S$ is perturbative for the Type $IIB$ string, this $D=4$
triality implies $S$-duality for the heterotic string and thus fills
a gap left by $D=6$ duality. For all
three strings the total symmetry is $SL(2,\BbbZ)_S \times
O(6,22;\BbbZ)_{TU}$.  The $O(6,22;\BbbZ)$ is {\it perturbative} for the
heterotic string but contains the conjectured {\it non-perturbative}
$SL(2,\BbbZ)_X$, where $X$ is the complex scalar of the $D=10$ Type
$IIB$ string.  Thus four-dimensional triality also provides a
(post-compactification) justification for this conjecture. We interpret
the $N=4$ Bogomol'nyi spectrum from all three points of
view.  In particular we generalize the Sen-Schwarz formula for short
multiplets to include intermediate multiplets also and discuss the
corresponding black hole spectrum both for the $N=4$ theory and for a
truncated $S$--$T$--$U$ symmetric $N=2$ theory.  Just as the first
two strings are described by the four-dimensional {\it elementary} and
{\it dual solitonic} solutions, so the third string is described by the
{\it stringy cosmic string} solution.  In three dimensions all three
strings are related by $O(8,24;\BbbZ)$ transformations.

\vfill
\leftline{July 1995}

\newpage

\section{Introduction}
\la{Introduction}

An interesting special case of string/string duality
\cite{Luloop,Khurifour,Lublack,Minasian,Duffclassical,Khuristring,
Duffstrong} is
provided by the $D=10$ heterotic string compactified to $D=6$ on $T^4$
which is related by strong/weak coupling to the $D=10$ Type $IIA$
string compactified to $D=6$ on $K3$ \cite{Hull,Witten,Senssd,Harvey}.
The dilaton
${\tilde \Phi}$, metric ${\tilde G}_{MN}$ and $2$-form ${\tilde
B}_{MN}$ of the Type $IIA$ theory are related to those of the heterotic
theory, $\Phi$, $G_{MN}$ and $B_{MN}$, by
\cite{Luloop,Lublack,Minasian,Khuristring,Duffstrong}
\begin{eqnarray}
{\tilde \Phi}&=&-\Phi\nonumber\\
{\tilde G}_{MN}&=&e^{-\Phi}G_{MN}\nonumber\\
{\tilde H}&=&e^{-\Phi}*H\ ,
\la{dual}
\end{eqnarray}
where $M=0,\ldots,5$, $H=dB+\cdots$, ${\tilde H}=d{\tilde B}$ and $*$
denotes the Hodge dual.  This ensures that the roles of $3$-form field
equations and Bianchi identities in one version of the corresponding
supergravity theory are interchanged in the other.

After further toroidal compactification to $D=4$ this automatically
accounts for the conjectured strong/weak coupling $SL(2,\BbbZ)$
duality in the resulting $D=4$, $N=4$ Type $IIA$ string and hence
for the $N=4$ Yang-Mills theories obtained by taking the global
limit \cite{Duffstrong}.
This is because $S$, the four-dimensional axion/dilaton
field, and $T$, the complex
Kahler form of the torus, are interchanged in going from the heterotic
to the Type $IIA$ theory. Moreover, while the
electric field strengths of the Kaluza-Klein gauge fields arising from
$G_{MN}$ are the same in both pictures, those of the ``winding" gauge fields
arising from $B_{MN}$ in the heterotic theory are replaced by their magnetic
duals in the Type $IIA$ theory.  Thus the strong/weak coupling duality
of the Type $IIA$ string is just the target-space $SL(2,\BbbZ)_T$ of
the heterotic string.

However, the target space symmetry of the heterotic theory also
contains an $SL(2,\BbbZ)_U$ that acts on $U$, the complex structure of
the torus%
\footnote{In this paper, the phrase {\it U-duality} will be taken to
mean $SL(2,\BbbZ)_U$ called $SL(2,\BbbZ)_O$ in \cite{Duffstrong}.  This
should not be confused with the $U$-duality of \cite{Hull} where it was
taken to mean the conjectured $E_7$ duality \cite{Luduality} of the
toroidally compactified Type $II$ string.}.
This suggests that, in addition to these $S$ and $T$ strings there
ought to be a third {\it $U$-string} whose axion/dilaton field is $U$
and whose strong/weak coupling duality is $SL(2,\BbbZ)_U$.  From a
$D=6$ perspective, this seems strange since, instead of (\ref{dual}),
we now interchange $G_{45}$ and $B_{45}$. Moreover, of the two electric
field strengths which become magnetic, one is a winding gauge field and
the other is Kaluza-Klein! So such a duality has no $D=6$ Lorentz
invariant meaning.  In fact, this $U$ string is a Type $IIB$ string, a
result which may also be understood from the point of view of
mirror%
\footnote{We are grateful to Xenia De La Ossa and Jan Luis for pointing
out that $T$--$U$ interchange is a mirror symmetry.}
symmetry:  interchanging the roles of Kahler form and complex structure
(which is equivalent to inverting the radius of one of the two circles)
is a symmetry of the heterotic string but takes Type $IIA$ into Type
$IIB$ \cite{Dai,Dine}.  In summary, if we denote the heterotic, $IIA$
and $IIB$ strings by $H,A,B$ respectively and the axion/dilaton,
complex Kahler form and complex structure by the triple $XYZ$ then we
have a triality between the $S$-string ($H_{STU}=H_{SUT}$), the
$T$-string ($B_{TUS}=A_{TSU}$) and the $U$-string ($A_{UST}=B_{UTS}$)
as illustrated in Fig. (1).

\begin{figure}
\centerline{\epsfbox{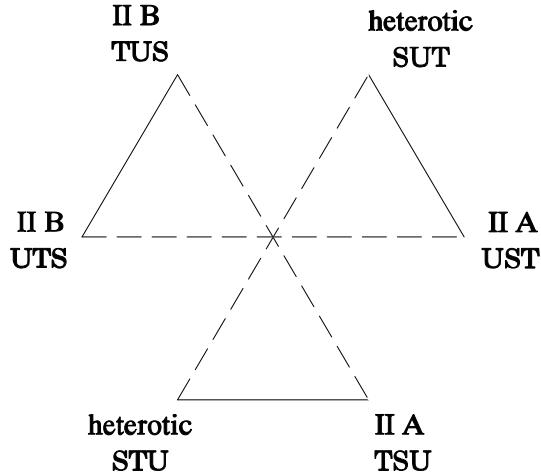}}
\medskip
\caption{String/string/string triality.  The solid lines
correspond to string/string dualities and the dashed lines
represent mirror transformations.}
\end{figure}

The field theory limits of the heterotic string on $T^4$, the Type
$IIA$ string on $K3$ and the Type $IIB$ string on $K3$ are described by
certain $N=2,D=6$ supergravity theories described in section
(\ref{sec:N2D6}).  As discussed in detail in section (\ref{sec:N4D4}),
each string in $D=4$ will then exhibit the same total symmetry
\be
SL(2,\BbbZ)_S \times O(6,22;\BbbZ)_{TU} \supset
SL(2,\BbbZ)_S \times SL(2,\BbbZ)_T \times SL(2,\BbbZ)_U\ ,
\ee
with the $28$ gauge field strengths and their duals
transforming as a $(2,28)$,
albeit with different interpretations for the three $SL(2,\BbbZ)$
factors.  Note that there is a discrete symmetry under $T$--$U$
interchange, but there is no such $U$--$S$ or $S$--$T$ symmetry.  As
discussed in \cite{Duffstrong}, it is the degrees of freedom associated
with going from $10$ to $6$ which are responsible for this lack of
$S$--$T$--$U$ democracy.  This will also be reflected in the
Bogomol'nyi spectrum of electric and magnetic states that belong to the
short and intermediate $N=4$ supermultiplets.  It is therefore
instructive to consider first the simpler situation where these modes
are truncated out.  This we do first in section (\ref{sec:N1D6}) by
truncating the $N=2,D=6$ supergravities to $N=1,D=6$ and then in
section (\ref{sec:N2D4}) by reducing these supergravities to $D=4$.  We
write down the action which describes the low energy limit of the
$S$-string; it exhibits an off-shell (perturbative) $SL(2,\BbbZ)_T
\times SL(2,\BbbZ)_U$ symmetry%
\footnote{The classical supergravities will in fact display continuous
symmetries such as $SL(2,R)$, but since these will be broken by quantum
corrections to discrete symmetries such as $SL(2,\BbbZ)$, we shall from
now on refer only to these.}
and an on-shell (non-perturbative) $SL(2,\BbbZ)_S$. Similarly, the
$T$-string action has an off-shell $SL(2,\BbbZ)_U \times SL(2,\BbbZ)_S$
and an on-shell $SL(2,\BbbZ)_T$, while the $U$-string action has an
off-shell $SL(2,\BbbZ)_S \times SL(2,\BbbZ)_T$ and an on-shell
$SL(2,\BbbZ)_U$.  Aside from the pedagogical usefulness of this
$S$--$T$--$U$ symmetric truncation, which describes just $4$ of the
$28$ gauge fields, it will turn out that this theory and the resulting
$S$--$T$--$U$ symmetric Bogomol'nyi spectrum, discussed in section
(\ref{sec:N2Sol}), will find application in $N=2$ theories whose
Bogomol'nyi spectrum includes multiplets which were both short and
intermediate from the $N=4$ point of view. In particular we discuss the
extreme black hole spectrum \cite{Rahmfeld1,Senblacktorus,Kalloshpeet,Cvetic}.

In section (\ref{sec:N2Sol}) we provide a soliton interpretation of the
three strings. We identify the $S$-string with the {\it elementary
string} solution of \cite{Dabholkar}, the $T$-string with the {\it dual
solitonic string} solution of \cite{Khurifour} and the $U$-string with
(a limit of) the {\it stringy cosmic string} solution of \cite{Greene}.
In $D=3$ dimensions, all three strings are related by $O(4,4;\BbbZ)$
transformations.

In sections (\ref{sec:N2D6}), (\ref{sec:N4D4}), (\ref{sec:N4Bog}) and
(\ref{sec:N4Sol}) we repeat the exercise of sections (\ref{sec:N1D6}),
(\ref{sec:N2D4}), (\ref{sec:N2Bog}) and (\ref{sec:N2Sol}), now
including the full set of states.  Section (\ref{sec:N2D6}) describes
the three $N=2$, $D=6$ supergravities: the actions in the heterotic and
Type $IIA$ cases (together with a duality dictionary relating the two
sets of fields) and the equations of motion in the case of Type $IIB$.
The compactification to $N=4$, $D=4$ of section (\ref{sec:N4D4})
reveals one or two surprises: although the $S$-string action has an
off-shell $O(6,22;\BbbZ)$ which continues to contain $SL(2,\BbbZ)_T
\times SL(2,\BbbZ)_U$, the $T$-string action has only an off-shell
$SL(2,\BbbZ)_U \times O(3,19;\BbbZ)$ which does not contain $SL(2,\BbbZ)_S$.
Similarly, the $U$-string action has only an $SL(2,\BbbZ)_T \times
O(3,19;\BbbZ)$ which does not contain $SL(2,\BbbZ)_S$.  In short, none of
the actions is $SL(2,\BbbZ)_S$ invariant! This lack of off-shell
$SL(2,\BbbZ)_S$ in the Type $II$ actions can be traced to the presence
of the extra $24$ gauge fields which arise from the R-R sector of Type
$II$ strings: $S$-duality in the heterotic picture acts as an on-shell
electric/magnetic transformation on all $28$ gauge fields and continues
to be an on-shell transformation on the $24$ which remain unchanged
under the string/string/string triality%
\footnote{The absence of a $R\rightarrow 1/R$ $T$-duality symmetry of
the Type $II$ supergravity action in $D=9$ has been noted in
\cite{Bergshoeffduality}.}.

At first sight, this seems disastrous for deriving the strong/weak
coupling duality of the heterotic string from target space duality of
the Type $II$ string.  The whole point was to explain a {\it
non-perturbative} symmetry of one string as a {\it perturbative}
symmetry of another \cite{Duffstrong}.  Fortunately, all is not lost:
although $SL(2,\BbbZ)_S$ is not an off-shell symmetry of the Type $II$
supergravity
actions, it is still a symmetry of the Type $II$ string
theories.  To see this we first note that $D=6$ general covariance is a
perturbative symmetry of the Type $IIB$ string and therefore that the
$D=4$ Type $IIB$ strings must have a perturbative $SL(2,\BbbZ)$ acting
on the complex structure of the compactifying torus.  Secondly we note
that for both Type $IIB$ theories, $B_{TUS}$ and $B_{UTS}$, $S$ is the
complex structure field.  Thus the $T$ string has $SL(2,\BbbZ)_U
\times SL(2,\BbbZ)_S$ and the $U$ string has
$SL(2,\BbbZ)_S \times SL(2,\BbbZ)_T$ as
required%
\footnote{We are grateful to Ashoke Sen for discussions on
these issues.}.
In this sense, four-dimensional string/string/string triality fills
a gap left by six-dimensional string/string duality: although duality
satisfactorily
explains the strong/weak coupling duality of the $D=4$ Type $IIA$
string in terms of the target space duality of the heterotic string,
the converse requires the Type $IIB$ ingredient.

Note that all of the three $SL(2,\BbbZ)_{(S,T,U)}$ take NS-NS states
into NS-NS states and that none can be identified with the conjectured
non-perturbative $SL(2,\BbbZ)_X$, where $X$ is the complex scalar of
the Type $IIB$ theory in $D=10$, which transforms NS-NS into R-R
\cite{Callan2,Hull,Witten}.  However, this $SL(2,\BbbZ)_X$ is a
subgroup of $O(6,22;\BbbZ )$.  Since this is a perturbative target space
symmetry of the heterotic string, the conjecture follows automatically
from the $D=4$ string/string/string triality hypothesis. Thus
we can say that evidence for this triality is evidence
not only for the electric/magnetic duality of all three $D=4$ strings
but also for the $SL(2,\BbbZ)_X$ of the $D=10$ Type $IIB$ string and
hence for {\it all} the conjectured non-perturbative symmetries of
string theory%
\footnote{One might object that in one case we have a
pre-compactification explanation but in the other only a
post-compactification explanation.  However, having established
$SL(2,\BbbZ)_X$ in the compactified version, its presence in the
uncompactified version then follows by blowing up the extra dimensions
keeping fixed the complex $X$ field.  We are grateful to Ashoke Sen
for this observation.}.

In section (\ref{sec:N4Bog}) we describe the $N=4,D=4$ Bogomol'nyi
spectrum. We generalize the heterotic string formula of Schwarz and
Sen, deriving the two $SL(2,\BbbZ)_S \times O(6,22;\BbbZ)_{TU}$
invariant central charges $Z_1$ and $Z_2$. This enables us to describe
the intermediate multiplets as well as the short ones, and once again we see
how the extreme black holes fit into this classification.

Section (\ref{sec:N4Sol}) generalizes (as far as is possible) the
soliton interpretation of section (\ref{sec:N2Sol}).  But as discussed in
\cite{Duffstrong}, including the extra degrees of freedom in going from
$10$ to $4$ causes problems in identifying the soliton zero modes.
Although it is straightforward to find the heterotic string as a
soliton of Type $II$, the converse is more problematical
\cite{Senssd,Harvey}.  In three dimensions, the $O(4,4;\BbbZ)$
generalizes to $O(8,24;\BbbZ)$
\cite{Luduality,Marcus,Rahmfeld1,Sen7,Rahmfeld2}.

Four-dimensional {\it string/string/string triality} was announced by
one of us (MJD) at the PASCOS 95 conference in Baltimore and at the
SUSY 95 conference in Paris \cite{Duffelectric}. Related results have
been obtained independently by Aspinwall and Morrison
\cite{Aspinwall2}.

\section{$N=1$ supergravity in $D=6$}
\label{sec:N1D6}

As a good guide to the kind of dualities one might expect in string
theory, it pays to look first at the corresponding supergravity
theories.  We therefore review some properties of $D=6$ supergravity
\cite{Salam}.  The theories of interest, which follow
either from $T^4$ compactification of the $D=10$ heterotic string or
$K3$ compactification of Type $II$, will be $N=2$ supergravities in
$D=6$ which yields $N=4$ in $D=4$.  All these theories are non-minimal in
the sense that they contain additional $N=2$ gauge or matter multiplets.
Since such additional matter destroys the $S$--$T$--$U$ symmetry of the
four-dimensional string we begin by examining an $N=1$ subset common to all
the models of interest.  We return to the full $N=2$ theory in section
(\ref{sec:N2D6}).

In terms of six-dimensional $N=1$ representations, we focus on the
supergravity multiplet $(G_{MN},\Psi^{+A}{}_M,B^+{}_{MN})$ and the
self-dual tensor multiplet $(B^-{}_{MN},\chi^{+A},\Phi)$.  The index
$A=1,2$ labels the $\bf 2$ of $Sp(2)$ and both spinors are symplectic
Majorana-Weyl.  The $2$-forms $B^+{}_{MN}$ and $B^-{}_{MN}$ have
$3$-form field strengths that are self-dual or anti-self-dual,
respectively.  Only with the combination of one supergravity multiplet
and one self-dual tensor multiplet do we have a conventional Lagrangian
formulation.  In this case the bosonic fields correspond to the
graviton, antisymmetric tensor and dilaton of string theory.  This
simpler theory will not only serve as a warm-up exercise for
understanding the $N=4,D=4$ superstrings but is interesting in its own
right for understanding the $N=2,D=4$ strings.

There are three theories to consider, each with the same number of
physical degrees of freedom.  The first two theories arise from the
truncation of the non-chiral $N=2$ supergravity and are related by
duality:  the first has the usual $3$-form field strength $H$ and the
second has the dual field strength ${\tilde H}=e^{-\Phi}*H$.  The third
theory comes from the truncation of the chiral $N=2$ supergravity.
While the full chiral $N=2$ theory does not admit a covariant
Lagrangian, the $N=1$ truncation, involving the combination of the
supergravity and tensor multiplet given above, may be written in a
conventional form.  In anticipation of their future application, we
shall call these theories $H$, $A$ and $B$, respectively.

Denoting the $D=6$ spacetime indices by
$(M,N=0,...,5)$, the bosonic part of the usual action takes the
form
\begin{equation}
I_H=\frac{1}{2\kappa^2}\int d^6x \sqrt{-G}e^{-\Phi}\left[
R_G+G^{MN}\partial_M\Phi\partial_N\Phi
-\frac{1}{12}G^{MQ}G^{NR}G^{PS}H_{MNP}H_{QRS}\right]\ .
\la{H}
\end{equation}
$H$ is the curl of the $2$-form $B$
\be
H=dB
\la{b}
\ee
(at this point there is no Chern-Simons correction).  The metric
$G_{MN}$ is related to the canonical Einstein metric $G^c{}_{MN}$ by
\be
G_{MN}=e^{\Phi/2}G^c{}_{MN}\ .
\la{metric}
\ee
Similarly, the dual supergravity action is given by
\begin{equation}
I_A=\frac{1}{2\kappa^2}\int d^6x \sqrt {-\tilde G}e^{-\tilde\Phi}\left[
R_{\tilde G}+\tilde G^{MN}\partial_M\tilde{\Phi}\partial_N\tilde{\Phi}
-\frac{1}{12}\tilde G^{MQ}\tilde G^{NR}\tilde G^{PS}
\tilde H_{MNP}\tilde H_{QRS}\right]\ .
\la{A}
\end{equation}
$\tilde H$ is also the curl of a 2-form $\tilde B$
\be
\tilde H=d\tilde B\ .
\la{d}
\ee
The dual metric $\tilde G_{MN}$ is related to the canonical Einstein
metric by
\be
\tilde G_{MN}=e^{\tilde\Phi/2}G^c{}_{MN}\ .
\la{e}
\ee
The two supergravities are related by:
\begin{eqnarray}
\tilde\Phi&=&-\Phi\nonumber\\
\tilde G_{MN}&=&e^{-\Phi}G_{MN}\nonumber\\
\tilde H&=&e^{-\Phi}\, {\ast H}\ ,
\la{duality}
\end{eqnarray}
where $\ast$ denotes the Hodge dual. (Since the last equation is
conformally invariant, it is not necessary to specify which metric is
chosen in forming the dual.) This ensures that the roles of field
equations and Bianchi identities in the one version of supergravity are
interchanged in the other.  The combined field equations and Bianchi
identities therefore exhibit a discrete symmetry under interchange of
$\Phi\rightarrow -\Phi$, $G\to \tilde G$ and $H\rightarrow \tilde H$.

Finally, while the third theory is unrelated to the other two (at least
in $D=6$), at this level of truncation it has a bosonic action with a
form similar to that of $I_A$.  One subtlety is worth mentioning, however.
Since this model arises from a truncation of the compactified Type $IIB$
string which has a {\it complex} 3-form field strength in ten dimensions,
there is some ambiguity in the identification of the dilaton
$\Phitt$ and 3-form $\Htt$ of model B, given in
the action
\begin{equation}
I_B=\frac{1}{2\kappa^2}\int d^6x
\sqrt{-\Gtt} e^{-\Phitt}\left[R_{\Gtt}
+\Gtt^{MN}\partial_M\Phitt\partial_N\Phitt
-\frac{1}{12}\Gtt^{MQ}\Gtt^{NR}\Gtt^{PS} \Htt_{MNP}\Htt_{QRS}\right]\ .
\la{eq:B}
\end{equation}
In particular, the $SL(2,\BbbZ)_X$ symmetry of the Type $IIB$
supergravity will mix $\Htt$ with its counterpart.  Nevertheless, from
a stringy viewpoint, we may identify $e^{\Phitt}$ as the string loop
expansion parameter and $\Htt$ as the 3-form field strength arising
from the NS-NS sector of the string.  This provides a unique definition
of the truncated action, (\ref{eq:B}).  Note that there is no $D=6$
Lorentz invariant dictionary between the fields $(\Phitt, \Gtt,\Htt)$
and $(\Phi,G,H)$ or $(\tilde\Phi, \tilde G,\tilde H)$.

\section{The $S$-$U$-$T$ symmetric theory in $D=4$}
\label{sec:N2D4}

Now let us first consider the H theory, dimensionally
reduced to $D=4$.  The combination of the six-dimensional $N=1$ supergravity
and tensor multiplets reduce to give the $D=4$, $N=2$ graviton multiplet
with helicities $(\pm2,2(\pm{3\over2}),\pm1)$ and three vector multiplets
with helicities $(\pm1,2(\pm{1\over2}),2(0))$.  In order to make this
explicit, we use a standard decomposition of the six-dimensional
metric
\begin{equation}
G_{MN}=\pmatrix{g_{\mu\nu}+A_\mu^mA_\nu^nG_{mn}&A_\mu^mG_{mn}\cr
A_\nu^nG_{mn}&G_{mn}}\ ,
\label{metricred}
\end{equation}
where the spacetime indices are ${\mu,\nu}=0,1,2,3$ and the internal indices
are $m,n=1,2$.  The remaining two vectors arise from the reduced $B$ field
\begin{equation}
B_{MN}=\pmatrix{B_{\mu\nu}+\half(A_\mu^mB_{m\nu}+B_{\mu n}A_\nu^n)&
B_{\mu n}+A_\mu^mB_{mn}\cr
B_{m\nu}+B_{mn}A_\nu^n&B_{mn}}\ .
\end{equation}
Four of the six resulting scalars are moduli of the 2-torus.  We parametrize
the internal metric and $2$-form as
\be
G_{mn}=e^{\rho-\sigma}\left(
\begin{array}{cc}
e^{-2\rho}+c^2&-c\\
-c&1
\end{array}\right)\ ,
\ee
and
\be
B_{mn}=b\,\epsilon_{mn}\ .
\ee
The four-dimensional metric, given by $g_{\mu\nu}$, is
related to the four-dimensional canonical Einstein, $g^c_{\mu\nu}$, metric by
$g_{\mu\nu}=e^{\eta}g^c{}_{\mu\nu}$ where $\eta$ is the four-dimensional
shifted dilaton:
\be
e^{-\eta}=e^{-\Phi}\sqrt{\det G_{mn}}=e^{-(\Phi+\sigma)}\ .
\ee
Thus the remaining two scalars are the dilaton $\eta$ and axion $a$ where
the axion field $a$ is defined by
\be
\epsilon^{\mu\nu\rho\sigma}\partial_{\sigma}a=
\sqrt{-g}e^{-\eta}g^{\mu\sigma}g^{\nu\lambda}g^{\rho\tau}
H_{\sigma\lambda\tau}\ ,
\label{eq:Haxion}
\ee
where
\bea
H_{\sigma\lambda\tau}&=&3(\partial_{[\sigma}B_{\lambda\tau]}+\frac{1}{2}
A_{[\sigma}^m F_{\lambda\tau] m}+\frac{1}{2}B_{m [\sigma} F_{\lambda\tau]}^m)
\nn \\
F_{\lambda\tau}^m&=&\partial_\lambda A_\tau^m-\partial_\tau A_\lambda^m \\
F_{\lambda\tau m}&=&\partial_\lambda B_{m\tau}-\partial_\tau B_{m\lambda}\
  . \nn
\eea
[...] denotes antisymmetrization with weight one.

We may now combine the above six scalars into the complex axion/dilaton
field $S$, the complex Kahler form field $T$ and the complex structure
field $U$ according to
\begin{eqnarray}
S&=&S_1+iS_2=a+ie^{-\eta}\nonumber\\
T&=&T_1+iT_2=b+ie^{-\sigma}\nonumber\\
U&=&U_1+iU_2=c+ie^{-\rho}\ .
\end{eqnarray}
This complex parametrization allows for a natural transformation under the
various $SL(2,\BbbZ)$ symmetries.  The action of $SL(2,\BbbZ)_S$ is given by
\be
S \rightarrow \frac{aS+b}{cS+d}\ ,
\la{sl2zs}
\ee
where $a,b,c,d$ are integers satisfying $ad-bc=1$, with similar
expressions for $SL(2,\BbbZ)_T$ and $SL(2,\BbbZ)_U$.  Defining the
matrices ${\cal M}_S$, ${\cal M}_T$ and ${\cal M}_U$ via
\be
{\cal M}_S=\frac{1}{S_2}
\left(
\begin{array}{cc}
1 & S_1\\
S_1 & |S|^2
\end{array}
\right)\ ,
\label{eq:sl2mat}
\ee
the action of $SL(2,\BbbZ)_S$ now takes the form
\be
{\cal M}_S\rightarrow \omega_S{}^T{\cal M}_S\omega_S\ ,
\ee
where
\be
\omega_S=
\left(
\begin{array}{cc}
d& b\\
c & a
\end{array}
\right)\ ,
\ee
with similar expressions for ${\cal M}_T$ and ${\cal M}_U$.
We also define the $SL(2,\BbbZ)$ invariant tensors
\be
\epsilon_S=\epsilon_T=\epsilon_U=
\left(
\begin{array}{cc}
0& 1\\
-1 & 0
\end{array}
\right)\ .
\ee
The fundamental supergravity (\ref{H}) now becomes
\begin{eqnarray}
I_{STU}&=&\frac{1}{16\pi G}\int d^4x\sqrt{-g}e^{-\eta}\Bigl[
R_g + g^{\mu\nu}\partial_{\mu}\eta\partial_{\nu}\eta
-\frac{1}{12}g^{\mu\lambda}g^{\nu\tau}g^{\rho\sigma}
H_{\mu\nu\rho}H_{\lambda\tau\sigma}\nonumber\\
&&\kern9.3em
+\frac{1}{4}\Tr(\partial{\cal M}_T{}^{-1}\partial {\cal M}_T)
+\frac{1}{4}\Tr(\partial{\cal M}_U{}^{-1}\partial {\cal M}_U)\nonumber\\
&&\kern9.3em
-\frac{1}{4}{F_S}_{\mu\nu}{}^T({\cal M}_T \times {\cal M}_U){F_S}^{\mu\nu}
\Bigr]\ .
\la{S}
\end{eqnarray}
%

%
\begin{figure}
\epsfxsize=4in
\centerline{\epsfbox{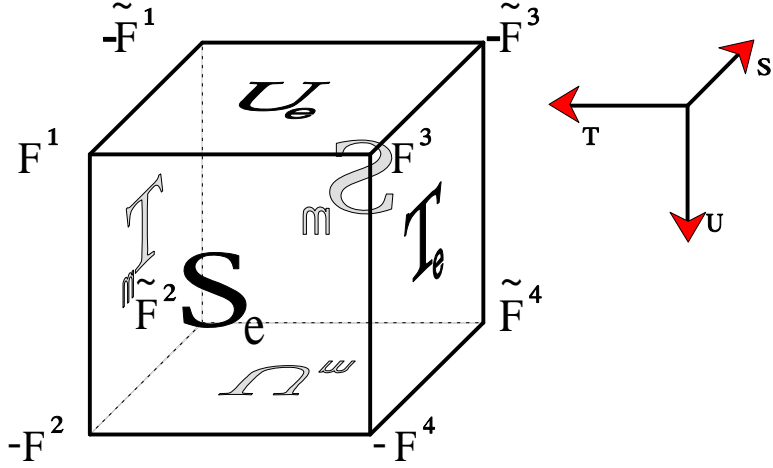}}
\medskip
\caption{The cube of triality. All field strengths are given in $S$-variables.}
\end{figure}
%

The four $U(1)$ gauge fields $A_S^a$ are given by
${A_S^1}_\mu=B_{4\mu}, \, {A_S^2}_\mu=B_{5\mu}, \, {A_S^3}_\mu=A_\mu^ 5,
\, {A_S^4}_\mu=-A_\mu^4$.  The three-form becomes
$H_{\mu\nu\rho}=3(\partial_{[\mu}B_{\nu\rho]}
-\half A_{S[\mu}{}^T (\epsilon_T\times\epsilon_U){F_S}_{\nu\rho]})$.
This action is
manifestly invariant under $T$-duality and $U$-duality, with
\be
{F_S}_{\mu\nu}\rightarrow
(\omega_T{}^{-1}\times\omega_U{}^{-1}){F_S}_{\mu\nu}\ , \, \qquad
{\cal M}_{T/U}\rightarrow \omega_{T/U}^T \, {\cal M}_{T/U} \, \omega_{T/U}\ ,
\ee
and with $\eta$, $g_{\mu\nu}$ and $B_{\mu\nu}$ inert.  Its equations of
motion and Bianchi identities (but not the action itself) are also
invariant under $S$-duality, with $T$ and $g^c{}_{\mu\nu}$ inert and
with
\be
\left(
\begin{array}{c}
{{F_S}}_{\mu\nu}{}^a\\
{\widetilde{F}_S}{}_{\mu\nu}{}^a
\end{array}
\right)
\rightarrow     \omega_S^{-1}
\left(
\begin{array}{c}
{{F_S}}_{\mu\nu}{}^a\\
{\widetilde{F}_S}{}_{\mu\nu}{}^a
\end{array}
\right)\ ,
\ee
where
\be
{\widetilde{F}_S}{}_{\mu\nu}{}^{a}=-S_2[({\cal M}_T{}^{-1} \times {\cal
M}_U{}^{-1})(\epsilon_T \times \epsilon_U)]^a{}_b  *
{F_S}_{\mu\nu}{}^{b}-S_1 {F_S}_{\mu\nu}{}^{a}  \ .
\ee
Thus $T$-duality transforms Kaluza-Klein electric charges
$({F_S}^3,{F_S}^4)$ into winding electric charges $({F_S}^1,{F_S}^2)$
(and Kaluza-Klein magnetic charges into winding magnetic charges),
$U$-duality transforms the Kaluza-Klein and winding electric charge of
one circle $({F_S}^3,{F_S}^2)$ into those of the other
$({F_S}^4,{F_S}^1)$ (and similarly for the magnetic charges) but
$S$-duality transforms Kaluza-Klein electric charge $({F_S}^3,{F_S}^4)$
into winding magnetic charge $({\tilde {F_S}}^3,{\tilde {F_S}}^4)$ (and
winding electric charge into Kaluza-Klein magnetic charge). In summary
we have $SL(2,\BbbZ)_T \times SL(2,\BbbZ)_U$ and $T \leftrightarrow U$
off-shell but $SL(2,\BbbZ)_S \times SL(2,\BbbZ)_T \times SL(2,\BbbZ)_U$
and an $S$--$T$--$U$ interchange on-shell.  The $S \leftrightarrow T$
part arises from the discrete on-shell symmetry $\Phi\rightarrow -\Phi$,
$G\to \tilde G$ and $H\rightarrow \tilde H$ in $D=6$.

Now consider the two actions obtained by cyclic permutation of the fields
$S,T,U$:
\begin{eqnarray}
I_{TUS}&=&\frac{1}{16\pi G}\int d^4x\sqrt{-{\tilde g}}e^{-\sigma}\Bigl[
R_{\tilde g} + {\tilde g}^{\mu\nu}\partial_{\mu}\sigma\partial_{\nu}\sigma
-\frac{1}{12}
{\tilde g}^{\mu\lambda}{\tilde g}^{\nu\tau}{\tilde g}^{\rho\sigma}
{\tilde H}_{\mu\nu\rho}{\tilde H}_{\lambda\tau\sigma}\nonumber\\
&&\kern9.3em
+\frac{1}{4}\Tr(\partial{\cal M}_U{}^{-1}\partial {\cal M}_U)
+\frac{1}{4}\Tr(\partial{\cal M}_S{}^{-1}\partial {\cal M}_S)\nonumber\\
&&\kern9.3em
-\frac{1}{4}{F_T}_{\mu\nu}{}^T({\cal M}_U \times {\cal M}_S){F_T}^{\mu\nu}
\Bigr]\ ,
\la{T}
\end{eqnarray}
and
\begin{eqnarray}
I_{UST}&=&\frac{1}{16\pi G}\int d^4x\sqrt{-\gtt}
e^{-\rho}\Bigl[R_{\gtt}+\gtt^{\mu\nu}\partial_\mu\rho\partial_\nu\rho
-\frac{1}{12}\gtt^{\mu\lambda} \gtt^{\nu\tau} \gtt^{\rho\sigma}
\Htt_{\mu\nu\rho}\Htt_{\lambda\tau\sigma}
\nonumber\\
&&\kern9.3em
+\frac{1}{4}\Tr(\partial{\cal M}_S{}^{-1}\partial {\cal M}_S)
+\frac{1}{4}\Tr(\partial{\cal M}_T{}^{-1}\partial {\cal M}_T)\nonumber\\
&&\kern9.3em
-\frac{1}{4}{F_U}_{\mu\nu}{}^T({\cal M}_S \times {\cal M}_T){F_U}^{\mu\nu}
\Bigr]\ .
\la{U}
\end{eqnarray}
The $D=6$ interpretation of these actions is as follows.  The action
$I_{TSU}=I_{TUS}$ is obtained by reducing the dual $A$ theory
(\ref{A}), where the four dimensional dual metric is given by ${\tilde
g}_{\mu\nu}=e^{\sigma}g^c{}_{\mu\nu}$ and the $3$-form field strength
${\tilde H}$ is related to the pseudoscalar field $b$ by
\be
\epsilon^{\mu\nu\rho\sigma}\partial_{\sigma}b=
\sqrt{-\tilde g}e^{-\sigma}\tilde g^{\mu\sigma}\tilde g^{\nu\lambda}
\tilde g^{\rho\tau} \tilde H_{\sigma\lambda\tau}\ .
\ee

However, since mirror symmetry interchanges $A$ and $B$ it also yields
the field equations obtained by reducing the field equations of the $B$
theory but with $S$ and $U$ interchanged.  Similarly, the action
$I_{UST}=I_{UTS}$ yields the field equations obtained by reducing the
$B$ theory, where the four dimensional metric is now given by
$\gtt_{\mu\nu} = e^{\rho}g^c{}_{\mu\nu}$ and the $3$-form field strength
$\Htt$ is related to the pseudoscalar field $c$ by
\be
\epsilon^{\mu\nu\rho\sigma}\partial_{\sigma}c= \sqrt{-\gtt}
e^{-\rho}{\gtt}^{\mu\sigma}{\gtt}^{\nu\lambda}
{\gtt}^{\rho\tau} {\Htt}_{\sigma\lambda\tau}\ .
\ee
Once again, however, by mirror symmetry this is equivalent to reducing
the $A$ theory with $S$ and $T$ interchanged.  The
relation between the field strengths ${F_S}$, ${F_T}$ and ${F_U}$ is
given in Table 1 and Figure 2. Figure 2 visualizes the
connection of all three strings. Each side of the cube
corresponds to electric or
magnetic $S$, $T$ or $U$ strings. Each dimension is related to one duality.
To get from one side to an adjacent one, two fields need to be dualized.
Mirror symmetry takes the cube into its mirror.

\begin{table}
$$
\begin{array}{ccccccc}
\rm axion/&\rm Kahler&\rm complex&\multispan4\hfil\rm gauge\ fields\hfil\\
\rm dilaton&\rm form&\rm structure&&&&\\
&&&&&&\\
S&T&U&{F_S}^3&-{F_S}^4&{F_S}^1&-{F_S}^2\\
S&U&T&{F_S}^3&F_S^1&-{F_S}^4&-{F_S}^2\\
U&S&T&{F_S}^3&{F_S}^1&-{\tilde {F_S}}^3&-\tilde{F_S}^1\\
U&T&S&{F_S}^3&-\tilde{F_S}^3&{F_S}^1&-\tilde{F_S}^1\\
T&U&S&{F_S}^3&-\tilde{F}_S^3&-F_S^4&\tilde{F}_S^4\\
T&S&U&{F_S}^3&-F_S^4&-\tilde{F}_S^3&{\tilde{F}_S}^4
\end{array}
$$
\label{table1}
\caption{Triality}
\end{table}

\section{The Bogomol'nyi Spectrum}
\label{sec:N2Bog}

It is now straightforward to write down an $S$--$U$--$T$ symmetric
Bogomol'nyi mass formula. Let us define electric and magnetic charge
vectors $\alpha_S^a$ and $\beta_S^a$ associated with the field strengths
${{F_S}}^a$ and ${\tilde {F_S}}^a$ in the standard way.
The electric and magnetic charges $Q_S^a$ and $P_S^a$ are
given by
\be {F_{S}}_{0r}^a\sim\frac{Q_S^a}{r^2} \, \qquad
*{F_{S}}_{0r}^a\sim\frac{P_S^a}{r^2}\ ,
\ee
giving rise to the charge vectors
\be
\pmatrix{\a_S^a\cr \b_S^a}=\pmatrix{  S_2^{(0)} {\cal M}_T^{-1}
   \times {\cal M}_U^{-1} & S_1^{(0)} \e_T \times \e_U  \cr 0 &
-\e_T \times \e_U }^{ab} \pmatrix{Q_S^b \cr P_S^b}.
\ee
For our purpose it is useful to define a generalized charge vector
$\gamma^{a{\tilde a}
{\tilde{\tilde a}}}$ via
\be
\left(
\begin{array}{c}
\gamma^{111}\\
\gamma^{112}\\
\gamma^{121}\\
\gamma^{122}\\
\gamma^{211}\\
\gamma^{212}\\
\gamma^{221}\\
\gamma^{222}
\end{array}
\right)
=
\left(
\begin{array}{c}
-\beta_S^1\\
-\beta_S^2\\
-\beta_S^3\\
-\beta_S^4\\
\alpha_S^1\\
\alpha_S^2\\
\alpha_S^3\\
\alpha_S^4
\end{array}
\right)\ ,
\ee
transforming as
\be
\gamma^{a{\tilde a}{\tilde {\tilde a}}}\rightarrow
\omega_S{}^a{}_b
\omega_T{}^{\tilde a}{}_{\tilde b}
\omega_U{}^{\tilde {\tilde a}}{}_{\tilde {\tilde b}}
\gamma^{b{\tilde b}{\tilde {\tilde b}}}\ .
\ee
Then the mass formula is
\begin{equation}
m^2=\frac{1}{16}\gamma^T({\cal M}_S{}^{-1}{\cal M}_T{}^{-1}{\cal M}_U{}^{-1}
		-{\cal M}_S{}^{-1}{\epsilon}_T{\epsilon}_U
		-{\epsilon}_S{\cal M}_T{}^{-1}{\epsilon}_U
		-{\epsilon}_S{\epsilon}_T{\cal M}_U{}^{-1})\gamma\ .
\la{us}
\end{equation}
Although all three theories have the same mass spectrum, there is
clearly a difference of interpretation with electrically charged
elementary states in one picture being solitonic monopole or dyon
states in the other. This agrees with the $N=2$ Bogomol'nyi formula of
Ceresole {\it et al.}\ \cite{Ceresole} and is a truncation of the generalized
$N=4$ mass formula derived from first principles
in section (\ref{sec:N4Bog}).  Note, however, that this is {\it not}
a truncation of the $N=4$ Bogomol'nyi formula of Schwarz and Sen
\cite{Schwarz1,Sen5}. In particular,
we note that
although both formulas have $SL(2,\BbbZ)_S \times SL(2,\BbbZ)_T \times
SL(2,\BbbZ)_U$, even the truncated
Schwarz-Sen formula (\ref{schwsen}) is only symmetric under $T$--$U$
interchange and not $S$--$T$--$U$. To understand this, we recall
that in $N=4$ supersymmetry, we have two central charges $Z_1$ and
$Z_2$.  There are three kinds of massive multiplets: short,
intermediate and long according as $(m=|Z_1|=|Z_2|)$, $(m=|Z_1|>|Z_2|)$
or $(m>|Z_1|,|Z_2|)$.  The Schwarz-Sen formula refers only to the short
multiplets. In $N=2$, however, we have only one central charge $Z$.
There are only short and long multiplets according as $m=|Z|$ or
$m>|Z|$.  States that were only intermediate in the $N=4$ theory may
thus become short in the truncation to $N=2$.

A nice example of this
phenomenon is provided by the extreme Reissner-Nordstrom black hole
(dilaton coupling $a=0$)
 which in string theory is dyonic with charge vectors
$\alpha=(1,0,0,-1)$ and $\beta=(0,-1,-1,0)$ \cite{Rahmfeld1}.  It
belongs to an intermediate multiplet in the $N=4$ theory and is
therefore absent from the Sen-Schwarz spectrum but belongs to a short
multiplet in the $N=2$ theory and appears in the spectrum (\ref{us}).
The two $N=4$ central charges are given in section (\ref{sec:N4Bog}).
Since we have identified the Reissner-Nordstrom black hole in the
$N=2$ spectrum, it is natural to ask which other black holes satisfy
(\ref{us}). Besides $a=0$, the supersymmetric
dilaton coupling parameters are $a=\sqrt{3},1,1/\sqrt{3}$
\cite{Khurinew,Rahmfeld1,Horowitz2,Khuriscatter}.  It turns out that all of the
corresponding states indeed satisfy the Bogomol'nyi bound and therefore
preserve 1/2 of the supersymmetries in the $N=2$ theory.
The $a=\sqrt{3}$ black hole has charge vectors $\a=(1,0,0,0)$,
$\b=(0,0,0,0)$. To cut a long story
short we set all the VEV's to zero and find its mass to be (in our units)
$m=1/4$, according to
\be
m^2=\frac{Q^2}{4(1+a^2)}\ ,
\ee
where $Q$ is the charge of the effective field strength.
Mass and charges are obviously related by (\ref{us}).
The mass of the
electrically charged $a=1$ black hole with $\a=(1,0,0,-1)$ is
$m=1/2$ \cite{Rahmfeld1}
which agrees also with (\ref{us}). Like the $a=\sqrt{3}$ black hole, this
solution
is elementary for the S-string, but it is dyonic for the $T$- and $U$-strings.
Further dynamical
evidence for the identification of $a=\sqrt{3}$ and $a=1$ black holes
with elementary heterotic
$N_L=1$ and $N_L>1$ string states \cite{Rahmfeld1} has recently
been given in \cite{Khuriscatter}.
Finally, the $a=1/\sqrt{3}$ black hole is dyonic in all pictures.
Its charge vectors are $\a=(1,0,0,-1)$ and $\b=(0,-1,0,0)$. The mass
is $m=3/4$ which can be verified by truncating the supergravity theory to
one effective field strength $\sqrt{3}F=F_S^1=-F_S^4=\tilde{F}_S^2$ along the
lines of \cite{Rahmfeld1}. A quick comparison with the Bogomol'nyi formula
proves that the $a=1/\sqrt{3}$ black hole preserves indeed 1/2
of the supersymmetries in $N=2$. As described in \cite{Rahmfeld1}, the
mass and charge assignments of the  $a=\sqrt{3},1,1/\sqrt{3}$ and $0$
black holes are compatible with their interpretations as $1,2,3$ and
$4$-particle bound states with zero binding energy.

\section{Soliton Interpretation}
\label{sec:N2Sol}

Four-dimensional string/string/string/triality suggests that it ought
to be possible to describe the $S$-string, $T$-string and $U$-string as
elementary and solitonic solutions directly in four dimensions. This is
indeed the case.  The $H$ action (\ref{H}) admits as an elementary
solution the $S$-string
\begin{eqnarray}
ds^2&=&e^{\eta}(-d\tau^2+d\sigma^2)+dzd\bar{z}\nonumber\\
S&=&a+ie^{-\eta}=\frac{1}{2\pi i}\ln\frac{r}{r_0}\ ,
\la{string1}
\end{eqnarray}
where $z=x_2+ix_3$ corresponds to the transverse directions and
$r=|z|$.  It also admits as a soliton solution the dual $T$-string
\begin{eqnarray}
ds^2&=&-d\tau^2+d\sigma^2+e^{-\sigma}dzd\bar{z}\nonumber\\
T&=&b+ie^{-\sigma}=\frac{1}{2\pi i}\ln\frac{r}{r_0}\ .
\la{string2}
\end{eqnarray}
Furthermore, it admits as a soliton solution the $U$-string
\begin{eqnarray}
ds^2&=&-d\tau^2+d\sigma^2+e^{-\rho}dzd\bar{z}\nonumber\\
U&=&c+ie^{-\rho}=\frac{1}{2\pi i}\ln\frac{r}{r_0}\ .
\la{string3}
\end{eqnarray}
We recognize the $S$-string as the {\it elementary string} solution of
\cite{Dabholkar} and the $T$-string as the {\it dual string} solution
of \cite{Khurifour} but the $U$-string is given by a limit of the {\it
stringy cosmic string} of \cite{Greene} where the fields $\rho$ and $c$
are simply given by the internal metric
\be
\sqrt{G}G^{-1}={\cal M}_U=e^{\rho}
\left(
\begin{array}{cc}
1 & c\\
c & c^2+e^{-2\rho}
\end{array}
\right)\ .
\ee
Consequently, the $U$-string is a solution of pure gravity in $D=6$ as
discussed in \cite{Greene}.

It follows that the $A$ action (\ref{A}) admits the $T$-string as
the elementary solution and the $S$- and $U$-strings as the solitonic
solutions and that the $B$ action (\ref{eq:B}) admits the $U$-string as
the elementary solution and the $T$- and $S$-strings as the solitonic
solutions.  Note that we may generate new $S$, $T$- and $U$-string
solutions by making $SL(2,\BbbZ)_S$ transformations on (\ref{string1}),
$SL(2,\BbbZ)_{T}$ transformations on (\ref{string2}) and
$SL(2,\BbbZ)_{U}$ transformations on (\ref{string3}).  So there is
really an $SL(2,\BbbZ)$ family of solutions for each string.  Once again,
all this is consistent with string/string/string triality.

The fundamental string solution given in (\ref{string1}) corresponds to
the case where all four gauge fields $(F_S{}^1,F_S{}^2,F_S{}^3,F_S{}^4)$
have been set to zero.  But as described in \cite{Sen2} a more general
solution with non-vanishing gauge fields may be generated by making
$O(3,3)$ transformations on the neutral solution.  Such deformations
are possible since the original solution is independent of $x^0$ as
well as $x^4$ and $x^5$. However, since we want to keep the asymptotic
values of the field configurations fixed, this leaves us with an
$O(2,1) \times O(2,1)$ subgroup. Not every element of this subgroup
generates a new solution; there is an $O(2) \times O(2)$ subgroup
that leaves the solution invariant. Thus the number of independent
deformations is given by the dimension of the coset space $O(2,1)
\times O(2,1)/O(2) \times O(2)$ which is equal to four, corresponding
to the four electric charges of $U(1)^4$. Exactly analogous statements
now apply to the $T$-string (\ref{string3}) and $U$-string
(\ref{string3}) solutions.

All of the above transformations take each string into itself. We now
consider transformations that map one string into another.  If we
compactify the $H$ action (\ref{H}) to {\it three} dimensions on $T^3$
the on-shell $SL(2,\BbbZ)_S$ will combine with the off-shell
$O(3,3;\BbbZ)$ target space duality to form an on-shell
$O(4,4;\BbbZ)$.  Similar remarks apply to the $A$ and $B$ actions. It
follows that all three strings are mapped into one another by
$O(4,4;\BbbZ)$ transformations. That the {\it stringy cosmic string}
was related to the {\it elementary} string in this way was pointed out
in \cite{Sen7}; that the {\it dual string} was also related in this way
was pointed out in \cite{Rahmfeld2}.

\section{$N=2$ supergravity in $D=6$}
\label{sec:N2D6}

The preceding discussion has shown an interesting triality structure of the
$H$, $A$ and $B$ theories when compactified to four dimensions.  However,
until now we have omitted the additional $D=6$ matter and/or gauge fields
present in all models.  In this section we examine the full $D=6$, $N=2$
theories, and in the next section we incorporate the additional fields into
string/string/string triality.

We begin by focusing on the heterotic string compactified on a generic torus
to $D=6$ \cite{Narain,Narain2}.
The low-energy limit of this theory is described by a non-chiral
$N=2$ supergravity with one graviton multiplet and 20 Yang-Mills multiplets.
The bosonic action is given by
\begin{eqnarray}
I_H&=&\frac{1}{2\kappa^2}\int d^6x \sqrt{-G}e^{-\Phi}\Bigl[
R_G+G^{MN}\partial_M\Phi\partial_N\Phi
-\frac{1}{12}G^{MQ}G^{NR}G^{PS}H_{MNP}H_{QRS}\nonumber\\
&&\kern4em
+\frac{1}{8}G^{MN}\Tr(\partial_MML\partial_NML)
-\frac{1}{4}G^{MP}G^{NQ}F_{MN}{}^a(LML)_{ab}F_{PQ}{}^b\Bigr]\ ,
\la{eq:Hfull}
\end{eqnarray}
where $A_M{}^a$ are $24$ abelian gauge fields and
$H_{MNP}= 3(\partial_{[M}B_{NP]}+\half A_{[M}{}^aL_{ab}F_{NP]}{}^b)$.
The 80 scalars parametrize an $O(4,20)/O(4)\times O(20)$ coset and
are combined into the symmetric $24\times24$ dimensional matrix $M$
satisfying $MLM=L$ where $L$ is the invariant metric on $O(4,20)$:
\be
L=\pmatrix{0&I_4&0\cr I_4&0&0 \cr 0&0&-I_{16}}\ .
\la{L}
\ee
The action is invariant under the $O(4,20;\BbbZ)$ target space duality
transformations $M\rightarrow\Omega M\Omega^T$,
$A_{\mu}{}^a\rightarrow\Omega^{a}{}_{b}A_{\mu}{}^b$,
$G_{\mu\nu}\rightarrow G_{\mu\nu}$, $B_{\mu\nu}\rightarrow B_{\mu\nu}$,
$\Phi\rightarrow\Phi$, where $\Omega$ is an $O(4,20;\BbbZ)$ matrix satisfying
$\Omega^TL\Omega=L$.
The full $I_H$ action is invariant under non-chiral six-dimensional
$N=2$ supersymmetry transformations.  For convenience in writing down
fermionic equations, we use an underlying $D=10$ notation where the four
$D=6$ symplectic Majorana-Weyl spinors of the $N=2$ theory may be combined
into a ten-dimensional Majorana-Weyl spinor $\epsilon$.  Since we will
need the supersymmetry transformations of the gravitino and dilatino when
deriving the Bogomol'nyi mass bound, we list them here:
\begin{eqnarray}
\delta\psi_M&=&\left[\nabla_M-{1\over8}H_{MNP}\Gamma^{NP}
+{1\over2\sqrt{2}} (VLF)_{R\,MN}^{\overline{a}}\Gamma^N\Gamma^{\overline{a}}
-{1\over4}(\partial_M^{\vphantom{-1}} V_R^{\vphantom{-1}}
V^{-1}_R)^{\overline{a}}{}_{\overline{b}}
\Gamma^{\overline{a}\overline{b}}\right]\epsilon\nonumber\\
\delta\lambda&=&-{1\over4\sqrt{2}}\left[\Gamma^M\partial_M\Phi
-{1\over6}H_{MNP}\Gamma^{MNP}+{1\over2\sqrt{2}}(VLF)_{R\,MN}^{\overline{a}}
\Gamma^{MN} \Gamma^{\overline{a}}\right]\epsilon\ ,
\label{eq:hetsusy}
\end{eqnarray}
where the Dirac matrices may be given a ten-dimensional interpretation,
$\Gamma^{(10)}=\{\Gamma^A,\Gamma^{\overline{a}}\}$, with six-dimensional
Dirac matrices $\Gamma_M=E_M^A \Gamma^A$ \cite{Romansf4}.

Turning to the Type $IIA$ string compactified on $K3$, we find an identical
massless spectrum, corresponding to one $N=2$ supergravity multiplet coupled
to 20 $N=2$ Yang-Mills multiplets \cite{Nilsson2}.  This time the action
is given by
\begin{eqnarray}
I_A&=&\frac{1}{2\kappa^2}\int d^6x \sqrt {-\tilde G}e^{-{\tilde \Phi}}\Bigl[
R_{\tilde G}+\tilde G^{MN}\partial_M{\tilde \Phi}\partial_N{\tilde \Phi}
-\frac{1}{12}\tilde G^{MQ}\tilde G^{NR}\tilde G^{PS}
\tilde H_{MNP}\tilde H_{QRS}\nonumber\\
&&\kern6em
+\frac{1}{8}{\tilde G}^{MN}\Tr(\partial_M{\tilde M}L\partial_N{\tilde M}L)
-{1\over4}e^{\tilde\Phi}\tilde G^{MP}\tilde G^{NQ}\tilde F_{MN}^a(L\tilde
ML)_{ab}\tilde F_{PQ}{}^b\Bigr]\nonumber\\
&&-{1\over2\kappa^2}\int d^6x{1\over16}\epsilon^{MNPQRS}\tilde B_{MN}\tilde
F_{PQ}{}^aL_{ab}\tilde F_{RS}{}^b\ ,
\la{eq:Afull}
\end{eqnarray}
where now ${\tilde H}$ has no Chern-Simons corrections, ${\tilde
H}=d{\tilde B}$.  The action (\ref{eq:Afull}) has the same $O(4,20;\BbbZ)$
symmetry as (\ref{eq:Hfull}) \cite{Nilsson1}.
In particular, the matrix $\tilde M$ of scalars
satisfies the constraint $\tilde M L\tilde M=L$.

Under heterotic/Type $IIA$ duality we have the following dictionary
\cite{Duffstrong,Witten} relating the two sets of fields:
\begin{eqnarray}
{\tilde \Phi}&=&-\Phi\nonumber\\
{\tilde G}_{MN}&=&e^{-\Phi}G_{MN}\nonumber\\
{\tilde H}&=&e^{-\Phi}*H \\
\tilde A_M&=&A_M\nonumber\\
\tilde M&=&M\ .
\end{eqnarray}
This gives, in particular, the Type $IIA$ gravitino and dilatino
supersymmetry transformations
\begin{eqnarray}
\delta\tilde\psi_M&=&\Biggl[\tilde\nabla_M
-{1\over8}\tilde H_{MNP}\Gamma^{\hat7}\tilde\Gamma^{NP}\nonumber\\
&&\qquad-{1\over8\sqrt{2}} e^{\tilde\Phi/2}
(\tilde V L\tilde F)_{R\,NP}^{\overline{a}}
(\tilde\Gamma_M\tilde\Gamma^{NP}-4\delta_M{}^N\tilde\Gamma^P)
\Gamma^{\overline{a}} -{1\over4}
(\partial_M^{\vphantom{-1}} \tilde V_R^{\vphantom{-1}}
\tilde V^{-1}_R)^{\overline{a}}{}_{\overline{b}}
\Gamma^{\overline{a}\overline{b}}\Biggr]\tilde\epsilon\nonumber\\
\delta\tilde\lambda&=&{1\over4\sqrt{2}}\left[\tilde\Gamma^M\partial_M\tilde\Phi
+{1\over6}\tilde H_{MNP}\Gamma^{\hat7}\tilde\Gamma^{MNP}
-{1\over2\sqrt{2}}e^{\tilde\Phi/2}
(\tilde V L\tilde F)_{R\,MN}^{\overline{a}}
\tilde\Gamma^{MN} \Gamma^{\overline{a}}\right]\tilde\epsilon\ ,
\end{eqnarray}
where $\Gamma^{\hat7}$ is the six-dimensional chirality operator with
eigenvalues $\pm1$.
Actually, (\ref{eq:Afull}) is not quite the action obtained by
compactifying $IIA$ supergravity on $K3$ which really has only $23$
vectors and one $3$-form potential $A_{MNP}$ \cite{Duffliu}; we have taken the
liberty of dualizing the $3$-form. Note that before dualizing
the off-shell symmetry is only
$O(3,19;\BbbZ)$.

Finally we consider the compactification of the Type $IIB$ theory on $K3$
\cite{Townsendk3}.
Since this theory is chiral in ten dimensions, it yields the chiral $N=2$
theory in six dimensions with
1 supergravity and 21 tensor multiplets.  While this theory has
no covariant action, the equations of motion for the (anti)-self-dual
three-forms may be determined from the well-known properties of $K3$.
Details of this procedure are presented in the appendix.  The resulting
equations have an on-shell $O(5,21,\BbbZ)$ invariance with $5\times 21=105$
scalars parametrizing the coset $O(5,21)/O(5)\times O(21)$.  There are
$21+5=26$ chiral 3-forms, which we denote collectively as
$\Htt_3^{i\pm}$, satisfying the (anti)-self-duality condition
\begin{equation}
\Htt_3^{i\pm} = \etatt_{ij}
*\Htt_3^{j\pm}\ ,
\label{eq:IIBSD}
\end{equation}
with
\begin{equation}
\etatt=\pmatrix{-1\cr&1\cr&&-1\cr&&&1\cr&&&&\eta_{ij}}\ .
\end{equation}
We have written $\Htt_3^{i\pm}$ in a given order such that the
first 4 fields correspond to the self-dual and anti-self-dual components of
$H^{(1)}$ and $H^{(2)}$ (the ten-dimensional NS-NS and R-R 3-forms,
respectively).  The remaining 22 chiral 3-forms come from the
compactification of the ten-dimensional self-dual 5-form field strength on
$K3$.
These chiral 3-forms as a set satisfy 26 Bianchi identities/equations of
motion
\begin{equation}
d\cHtt_3^I=0\ ,
\end{equation}
where the two sets of 3-forms are related by a vierbein $\Vtt^i_J$
\begin{equation}
\cHtt_3=(\Ltt^{-1})(\Vtt^{-1}) \Htt_3^\pm\qquad
\Htt_3^\pm=\Vtt\Ltt \cHtt_3\ .
\end{equation}
The $O(5,21)$ matrix $\Ltt$ is given by
\begin{equation}
\Ltt=\pmatrix{-\sigma^1\cr&\sigma^1\cr&&d_{IJ}}\ ,
\label{eq:tildetildeL}
\end{equation}
and $\Vtt$ satisfies
\begin{equation}
\Vtt^{-1}=[\etatt\Vtt\Ltt]^T\ .
\end{equation}
The explicit form for $\Vtt$ is given in the appendix.
The equations of motion for the bosonic fields of model $B$ are given by
\cite{RomansSD}
\begin{eqnarray}
\Rtt_{MN}-\half \Gtt_{MN}\Rtt&=&
{1\over4}\Htt_{MPQ}^{i\pm}\Htt^{i\pm}{}_N{}^{PQ}\nonumber\\
&&+\Tr[\partial_M^{\vphantom{1}}\Vtt_R^{\vphantom{1}}\Vtt_L^{-1}
\partial_N^{\vphantom{1}}\Vtt_L^{\vphantom{1}}\Vtt_R^{-1}]
-{1\over2}\Gtt_{MN} \Tr[\partial_P^{\vphantom{1}}
\Vtt_R^{\vphantom{1}}\Vtt_L^{-1}
\partial^P\Vtt_L^{\vphantom{1}}\Vtt_R^{-1}]\nonumber\\
&&\kern-9.4em
\nabla^M(\partial_M^{\vphantom{1}}\Vtt_R^{\vphantom{1}}\Vtt_L^{-1})^{ij}
-(\partial_M^{\vphantom{1}}\Vtt_R^{\vphantom{1}}\Vtt^{-1}\etatt
\partial^M\Vtt\Vtt_L^{-1})^{ij}
= {1\over6} \Htt_{MNP}^{i+}\Htt^{j-\,MNP}\nonumber\\
\Htt_3^{i\pm}&=&\etatt_{ij}*\Htt_3^{j\pm}\nonumber\\
d\Htt_3^{i\pm}&=&(d\Vtt\Vtt^{-1})_{ij}\Htt_3^{j\pm}\ .
\end{eqnarray}
We note that the Type $IIB$ dilaton is included implicitly as one of the
scalars in $\Vtt$.  Thus the equations of motion
are written above in a canonical framework.
The supersymmetric variation of the canonical gravitino is
\begin{equation}
\delta\skew5\tilde{\tilde\psi}_M^a
=\left[\delta^a{}_b\nabla_M+{1\over4}\Htt_{MNP}^{i+}\Gamma^{NP}(T^i)^a{}_b
\right]\epsilon^b\ ,
\end{equation}
where the spinors $\epsilon^a$ are right-handed symplectic Majorana-Weyl
with $a$ labeling the $4$ of $Sp(4)\simeq SO(5)$.  The five self-dual
3-forms transform as a vector of $SO(5)$ and the matrices $T^i$ satisfy the
$SO(5)$ Clifford algebra $\{T^i,T^j\}=2\delta^{ij}$.  The (anti)self-duality
conditions are essential for the closure of the supersymmetry algebra
\cite{RomansSD}.

In order to gain a better understanding of model $B$, we may consider a
few special limits.  If we set the R-R moduli to zero, then the vierbein
(given in the appendix) decomposes as
\begin{equation}
\Vtt^i{}_J=
\left[\matrix{{1\over\sqrt{2}}e^{\Phitt}\cr
&{1\over\sqrt{2}}e^{\Phitt}\cr
&&{1\over\sqrt{2}}e^{\rho/2}\cr
&&&{1\over\sqrt{2}}e^{\rho/2}\cr
\vphantom{{1\over\sqrt{2}}}
&&&&\!I_{22}}\right]\times\left[\matrix{\vphantom{{1\over\sqrt{2}}}
-1&-e^{-\Phitt}\cr
\vphantom{{1\over\sqrt{2}}}1&-e^{-\Phitt}\cr
\vphantom{{1\over\sqrt{2}}}&&1&-e^{-\rho}-\half(b)^2&b^J\cr
\vphantom{{1\over\sqrt{2}}}&&-1&-e^{-\rho}+\half(b)^2&-b^J\cr
\vphantom{{1\over\sqrt{2}}}&&0&-O^i{}_Ib^I&\!\!O^i{}_Kd^{KJ}}\right]
\end{equation}
where $(b)^2=b^Ib^Jd_{IJ}$.  This shows explicitly the factorization into
the dilaton and the $O(4,20)$ moduli space of $K3$ with torsion.  Due to
the $D=10$ symmetry between $H^{(1)}$ and $H^{(2)}$, we may choose
to eliminate a different set of moduli, giving instead
\begin{equation}
\Vtt^i{}_J=
\left[\matrix{{1\over\sqrt{2}}e^{\Phitt}\cr
&{1\over\sqrt{2}}e^{\Phitt}\cr
&&{1\over\sqrt{2}}e^{\rho/2}\cr
&&&{1\over\sqrt{2}}e^{\rho/2}\cr
\vphantom{{1\over\sqrt{2}}}
&&&&\!I_{22}}\right]\times\left[\matrix{\vphantom{{1\over\sqrt{2}}}
-1&-e^{-\Phitt}-\half(b')^2&&&-b'^J\cr
\vphantom{{1\over\sqrt{2}}}1&-e^{-\Phitt}+\half(b')^2&&&b'^J\cr
\vphantom{{1\over\sqrt{2}}}&&1&-e^{-\rho}\cr
\vphantom{{1\over\sqrt{2}}}&&-1&-e^{-\rho}\cr
\vphantom{{1\over\sqrt{2}}}0&O^i{}_Ib'^I&&&\!\!O^i{}_Kd^{KJ}}\right]\!
\end{equation}
where now the $b'^I$ are R-R moduli arising from $H^{(2)}$.  This gives
a different decomposition of $O(5,21)$ into $O(1,1)\times O(4,20)$ and
hints at a symmetry under exchange of $\Phitt \leftrightarrow\rho$ where
$\rho$ is the $K3$ breathing mode.  In fact,
this is nothing but the underlying ten-dimensional $SL(2,\BbbZ)_X$
symmetry of the Type $IIB$ supergravity.  This may be made clear by
eliminating the torsion moduli, $b^I=b'^I=0$.  In this case the matrix
$\Mtt =\Vtt^T\Vtt$ may be written
\begin{equation}
\Mtt=\Omega\left[\matrix{{\cal M}_X\otimes{\cal M}_Y\cr
&H^I{}_Kd^{KJ}}\right]\Omega\ ,
\end{equation}
where $\Omega$ swaps entries 2 and 4.  The matrices ${\cal M}_X$ and ${\cal
M}_Y$ are $SL(2,\BbbZ)$ matrices defined according to (\ref{eq:sl2mat}) where
\begin{eqnarray}
X&=&-\ell+ie^{-(\Phitt-\rho)/2}\nonumber\\
Y&=&d+ie^{-(\Phitt+\rho)/2}
\end{eqnarray}
($d$ is the single modulus arising from the ten-dimensional 4-form
potential).  This shows a decomposition of $O(5,21)$ into $O(2,2)\times
O(3,19)$ with the last factor identified with the moduli of $K3$
surfaces of constant volume.  Since $\Phitt-\rho=\Phi^{(10)}$
is just the ten-dimensional dilaton, $X$ is exactly the field on which the
original $SL(2,\BbbZ)_X$ acts.

This last example may be further motivated by considering a truncated
version of model $B$ without self-dual fields.  The reduction of the
original ten-dimensional 3-forms gives
\begin{equation}
I_B^{H^{(i)}}={1\over4\kappa^2}\int\left[e^{-\Phitt}
H_3^{(1)}*H_3^{(1)}
+e^{-\rho}H_3^{(2')}*H_3^{(2')}\right]\ .
\label{eq:h1h2main}
\end{equation}
The $H^{(i)}$ are related to their counterparts in $D=10$ and are
explicitly
defined in the appendix. The on-shell symmetry of this version is the
$O(2,2;\BbbZ)$ subgroup of $O(5,21;\BbbZ)$
acting on the first four components.
One subgroup of this $O(2,2;\BbbZ)$ is the discussed $SL(2,\BbbZ)_X$.
Another interesting one is the $O(1,1;\BbbZ)\simeq \BbbZ_2$ acting on the
first two components. This transformation takes $H^{(1)}$ into
$e^{-\Phitt} *H^{(1)}$ and $\Phitt$ into
$-\Phitt$ and is therefore a strong/weak duality transformation
for the Type $IIB$ string. This transformation is precisely the one
transforming the $T$-string into the $U$-string.

\section{Reduction to $D=4$}
\label{sec:N4D4}

When models $H$, $A$ and $B$ are reduced to four dimensions, they all give
rise to $D=4$, $N=4$ supergravities coupled to 22 Yang-Mills multiplets.
{}From the heterotic point of view, it is straightforward to compactify the
six-dimensional theory, given by (\ref{eq:Hfull}), to four dimensions on a
two-torus.  The resulting bosonic action may be written
\begin{equation}
I_H^4={1\over16\pi G}\int d^4x\sqrt{-g}e^{-\eta}\left[R+(\partial\eta)^2
-{1\over12}{\cal H}_{\mu\nu\lambda}{}^2+{1\over8}\Tr(\partial \overline{M}
\overline{L}\partial\overline{M}\overline{L})-{1\over4}{\cal F}_{\mu\nu}{}^T
(\overline{L}\overline{M}\overline{L}){\cal F}_{\mu\nu}\right],
\end{equation}
where the four-dimensional variables are given by the standard dimensional
reduction techniques.  In particular, the 28 gauge fields ${\cal A}_\mu$
arise two from the metric, two from the antisymmetric tensor and 24 from the
gauge fields in six dimensions.  We group them together according to
\begin{equation}
{\cal A}_\mu=[A_\mu^i\quad \overline{B}_{i\mu}\quad \overline A_\mu]^T\ ,
\end{equation}
where
\begin{eqnarray}
\overline{A}_\mu&=&A_\mu-A_\mu^iA_i\nonumber\\
\overline{B}_{i\mu}&=&B_{i\mu}-A_\mu^jB_{ij}+\half\overline{A}_\mu^TLA_i\ .
\end{eqnarray}
Note that the six-dimensional gauge fields are denoted by $A_\mu$ whereas
the metric $U(1)$'s always carry an index $i=4,5$.  The scalars parametrize
an $O(6,22)/O(6)\times O(22)$ coset with metric
\begin{equation}
\overline{L}=\pmatrix{&I_2\cr I_2\cr&&L}
\end{equation}
and may be written in a vierbein form
\begin{equation}
\overline{V}
=\left[\matrix{{1\over\sqrt{2}}E^{-1}\cr&{1\over\sqrt{2}}E^{-1}\cr&&I_{24}}
\right]\times\left[\matrix{I_2&G+B-C&-A^T\cr I_2&-G+B-C&-A^T\cr
0&VLA&V}\right]\ ,
\end{equation}
where $C=\half A^TLA$ and $G$ and $B$ refer to the $4,5$ components of the
respective fields.  The 3-form ${\cal H}$ is dual to the axion as given by
$(\ref{eq:Haxion})$ and may be written ${\cal H}_{\mu\nu\lambda}=
3(\partial_{[\mu}\overline{B}_{\nu\lambda]}+\half{\cal A}_{[\mu}
\overline{L}{\cal F}_{\nu\lambda]})$ where
\begin{equation}
\overline{B}_{\mu\nu}=B_{\mu\nu}-A_\mu^iA_\nu^jB_{ij}-A_{[\mu}^i
(\overline{B}_{i\nu]}-A_i^TL\overline{A}_{\nu]})\ .
\end{equation}

It is of course no surprise that this theory has an explicit $O(6,22;\BbbZ)$
symmetry as expected from a direct compactification from ten dimensions on
$T^6$.  In fact, the above four dimensional action could have been written
directly without the extra step of compactifying to six dimensions.
However, for string/string/string triality, it is enlightning to see
explicitly the compactification from $D=6$ to $D=4$.  In particular, in the
absence of scalars $A_i$ originating from the six-dimensional gauge fields,
we find the simple split
\begin{equation}
\overline{V}={1\over\sqrt{2}}E^{-1}\left[\matrix{I_2&G+B\cr I_2&-G+B}\right]
\oplus V\ ,
\end{equation}
indicating the limit
\begin{equation}
{O(6,22)\over O(6)\times O(22)}\to
\left.{O(2,2)\over O(2)\times O(2)}\right|_{TU}
\times{O(4,20)\over O(4)\times O(20)}\ .
\end{equation}
Reduction of the Type $IIA$ theory on $T^2$ yields instead the
four-dimensional action
\begin{eqnarray}
I_A^4&=&{1\over16\pi G}\int d^4x\sqrt{-\tilde g}e^{-\tilde\eta}\biggl[
\tilde R+(\partial\tilde\eta)^2
-{1\over12} \tilde {\cal H}_{\mu\nu\lambda}{}^2
+{1\over4}(\Tr(\partial\tilde G^{-1}\partial\tilde G)
+\Tr(\partial\tilde B\tilde{G}^{-1}\partial\tilde B\tilde{G}^{-1}))\nonumber\\
&&\kern10.4em +{1\over8}\Tr(\partial \tilde{M}L \partial \tilde{M}L)
-{1\over4}(\tilde F_{\mu\nu}^i\tilde{G}_{ij}\tilde F_{\mu\nu}^j
+\tilde {\cal H}_{\mu i\nu}\tilde{G}^{ij}\tilde {\cal H}_{\mu j\nu})
\biggr]\nonumber\\
&&+{1\over16\pi G}\int d^4x\sqrt{-\tilde g}e^{-\tilde\sigma}\left[
-{1\over2}\Tr(\partial\tilde A^TL\tilde ML\partial\tilde A\tilde G^{-1})
-{1\over4}\tilde{\cal F}_{\mu\nu}^T (L\tilde{M}L)\tilde{\cal F}_{\mu\nu}
\right]\nonumber\\
&&+{1\over16\pi G}\int d^4x\biggl[
-{1\over8}\epsilon^{ij}\tilde B_{ij}
\tilde{\cal F}_{\mu\nu}^TL*\tilde{\cal F}_{\mu\nu}
-{1\over2}\epsilon^{ij}
(\tilde{\cal H}_{\mu i\nu}-\tilde B_{ik} \tilde F^k_{\mu\nu})
\tilde A_j^TL *(\tilde{\cal F}_{\mu\nu}-\half\tilde A_l \tilde F^l_{\mu\nu})
\nonumber\\
&&\kern6.55em -{1\over12}\epsilon^{\mu\nu\lambda\sigma}\epsilon^{ij}
\tilde{\cal H}_{\mu\nu\lambda}\tilde A_i^TL\partial_\sigma\tilde A_j
\biggr]\ .
\label{eq:IAfour}
\end{eqnarray}
As written, only $\tilde F_{\mu\nu}^i\equiv2\partial_{[\mu}^{\vphantom{i}}
\tilde A_{\nu]}^i$
are true field strengths ($\tilde{A}^{i}_{\mu}$ are the gauge fields arising
from the compactification of the metric $\tilde{G}$ as in (\ref{metricred})).
The other 2-forms, $\tilde{\cal H}_{\mu i\nu}$ and $\tilde{\cal F}_{\mu\nu}$,
are the shifted six-dimensional fields:
\begin{eqnarray}
\tilde{\cal H}_{\mu i\nu}&=& \tilde{H}_{\mu i\nu}
+2\tilde{A}^j_{[\mu}\partial_{\nu]}^{\vphantom{j}}\tilde{B}_{ij}
=2\partial_{[\mu}\overline{\tilde{B}}_{i\nu]}
+\tilde B_{ij}\tilde F_{\mu\nu}^j\nonumber\\
\tilde{\cal F}_{\mu\nu}&=& \tilde{F}_{\mu\nu}
\>\,+2\tilde{A}^j_{[\mu}\partial_{\nu]}^{\vphantom{j}}\tilde{A}_{j}
\ =2\partial_{[\mu}\overline{\tilde A}_{\nu]}
+\tilde A_i\tilde F_{\mu\nu}^i\ ,
\end{eqnarray}
where the four-dimensional gauge fields are
\begin{eqnarray}
\overline{\tilde B}_{i\mu}&=&\tilde B_{i\mu}-\tilde A_\mu^j\tilde B_{ij}
\nonumber\\
\overline{\tilde A}_\mu&=&\tilde A_\mu-\tilde A_\mu^i\tilde A_i\ .
\end{eqnarray}
%
$\tilde{\cal H}_{\mu\nu\lambda}$ is the three-form field strength with the
standard Bianchi identity arising from the metric and antisymmetric tensor
gauge fields:
\begin{eqnarray}
\tilde{\cal H}_{\mu\nu\lambda}&=&
3(\partial_{[\mu}\overline{\tilde B}_{\nu\lambda]}
+\tilde A_{[\mu}^i\partial_\nu^{\vphantom{i}}
\overline{\tilde B}_{i\lambda]}^{\vphantom{i}}
+\overline{\tilde B}_{i[\mu^{\vphantom{i}}}\partial_\nu^{\vphantom{i}}
\tilde A_{\lambda]}^i)\nonumber\\
\overline{\tilde B}_{\mu\nu}&=&\tilde B_{\mu\nu}
-\tilde A_{[\mu}^i\overline{\tilde B}_{i\nu]}^{\vphantom{i}}
-\tilde A_\mu^i\tilde A_\nu^j\tilde B_{ij}\ .
\end{eqnarray}

The duality map relating model $H_{STU}$ to model $A_{TSU}$ is given by
\begin{eqnarray}
&\hbox{metric}\hfil&\tilde g_{\mu\nu}=e^{\sigma-\eta}g_{\mu\nu}\nonumber\\
&\hbox{$U$ field}\hfil&\tilde G_{ij}=e^{\sigma-\eta} G_{ij}\nonumber\\
&\hbox{$S$--$T$ interchange}\hfil&\tilde\eta=\sigma\qquad
\tilde a=-\half\epsilon^{ij}B_{ij}\nonumber\\
&&\tilde\sigma=\eta\qquad \tilde B_{ij}=-\epsilon_{ij}a\nonumber\\
&\hbox{metric gauge fields}\quad&\tilde A_\mu^i=A_\mu^i\nonumber\\
&\hbox{$H$ gauge fields}\hfil&\tilde {\cal H}_{\mu i\nu}=e^{\sigma-\eta}
\epsilon_i{}^j*{\cal H}_{\mu j\nu}\nonumber\\
&\hbox{$D=6$ fields}\hfil&\overline{\tilde A}_\mu=\overline A_\mu
\qquad \tilde A_i=A_i\qquad \tilde M=M\ ,
\label{dualmap}
\end{eqnarray}
where $\eta(\tilde{\eta})$ and $\s(\tilde{\s})$ are the dilatons/T-moduli
of the relevant theories.

When reduced to four dimensions, model $B$ loses its chirality and now
admits a Lagrangian formulation.  Each six-dimensional three-form of
definite chirality reduces to a single $U(1)$ field strength and one
scalar.  Thus the 28 four-dimensional gauge fields come two from the
reduction of the metric and 26 from $\cHtt_3$.  Prior to
the imposition of the self-duality conditions, the latter field strengths
are given by
\begin{equation}
\Ftt^a_{i\, \mu\nu}=2\partial_{[\mu} \overline{\Btt^a}_{i\nu]} \qquad
\overline{\Btt^a}_{i\mu}=\Btt^a_{i\mu} -\Att_\mu^j\Btt^a_{ij}\ ,
\end{equation}
where $i=4,5$.  This gives a double counting which is eliminated by the
six-dimensional self-duality conditions, (\ref{eq:IIBSD}).  Thus
\begin{equation}
\cFtt_{i\,\mu\nu}^\pm=\epsilon_i{}^j \eta*\cFtt_{j\, \mu\nu}^\pm\ ,
\end{equation}
where
\begin{equation}
\cFtt_{i\,\mu\nu}^\pm=\Vtt(\Ftt_{i\, \mu\nu}+\Btt_{ij}\Ftt_{\mu\nu}^j)\ .
\end{equation}

Reduction of the six-dimensional 3-form field equations then give
\begin{equation}
\nabla_\mu\left[\Ltt\Mtt_{ab}
(\Ftt_i^{b\,\mu\nu}+\Btt^b_{ij}\Ftt^{j\,\mu\nu})
-\epsilon_i{}^j\Btt^a_{jk}*\Ftt^{k\,\mu\nu}\right]=0\ ,
\end{equation}
which is a set of $2\times26$ equations and should be viewed as a
combination of both Bianchi identities and equations of motion.  The
remaining equations of motion may similarly be reduced. We may then construct
a Type $IIB$ action which yields these equations of motion,
although there is some ambiguity in whether to choose
$p$-forms or their duals. The canonical choice is obtained by mirror
transformation
of the Type $IIA$ action, yielding the $B_{TUS}$ model. The duality map
relating $H_{SUT}$ to $A_{UST}$ is obtained by repeating (\ref{dualmap})
for the mirror-transformed heterotic string, and the $A_{UST}$ dilaton is then
$\rho$. The heterotic-Type $IIB$ dictionaries are then obtained by
performing mirror transformations on the Type $IIA$ strings.

{}From the conjectured six-dimensional heterotic/Type $IIA$ duality and
the connection between $IIA$ and $IIB$ via mirror symmetry it follows
that we have indeed a triality between all three strings in $D=4$;
beyond the simplified discussion of section (\ref{sec:N2D4}). However,
since $U$ and $T$
are embedded in the full $O(6,22;\BbbZ)$ whereas $S$ is not, the elegant
exchange symmetries $S/T$ and $S/U$ are destroyed. Note that the $A_{TSU}$
action (\ref{eq:IAfour}) has only $SL(2,\BbbZ)_U$
off-shell (besides the obvious $O(4,20;\BbbZ)$)
even though, as explained in the Introduction, the string
has also an $SL(2,\BbbZ)_S$. Similarly
the Type $B_{UTS}$ action has only $SL(2,\BbbZ)_T$
off-shell even though the Type $IIB$ string has also an
$SL(2,\BbbZ)_S$.
Consequently, none of the three actions is $SL(2,\BbbZ)_S$ invariant,
in contrast to the truncated $H,A,B$ actions discussed in section
(3). Since $SL(2,\BbbZ)_S$ is still a perturbative Type $IIB$ symmetry,
however, four-dimensional string/string/string triality still implies
the $S$-duality of the heterotic string.

\section{Bogomol'nyi Spectrum}
\label{sec:N4Bog}

We may derive the Bogomol'nyi mass bound in this theory by following a
Nester procedure \cite{Nester,Dabholkar,Harveyliu}.  Since masses are
defined with respect to a canonical metric, it is convenient to work in
canonical variables (which we denote by a caret).  From a supergravity
point of view, this mass bound
originates from the $N$-extended supersymmetry algebra with central charges
\cite{Wittenolive,Osborn}.
Thus we start by noting that, up to equations of motion, the supercharge
(parametrized by $\epsilon$) is given by
\begin{equation}
Q_{\epsilon}=\int\overline{\epsilon}\gamma^{\mu\nu\lambda}
\nabla_\nu\hat\psi_\lambda d\Sigma_\mu=\int\overline{\epsilon}
\gamma^{\mu\nu\lambda}\hat\psi_\lambda d\Sigma_{\mu\nu}\ .
\end{equation}
Therefore the anticommutator of two supercharges is
\begin{equation}
\{Q_\epsilon,Q_{\epsilon'}\}=\delta_\epsilon Q_{\epsilon'}
=\int N^{\mu\nu} d\Sigma_{\mu\nu}\ ,
\end{equation}
where $N^{\mu\nu}=\overline{\epsilon'}\gamma^{\mu\nu\lambda}
\delta_\epsilon\hat\psi_\lambda$ is a generalized Nester's form.

Just as the canonical Einstein metric is Weyl scaled by the dilaton
relative to the $\sigma$-model metric, the canonical gravitino is
shifted by the dilatino:
\begin{equation}
\hat\psi_\mu=e^{\eta/4}(\psi_\mu+\sqrt{2}\gamma_\mu\lambda)\ .
\end{equation}
Since the reduction of the six-dimensional supersymmetry transformations,
(\ref{eq:hetsusy}), gives
\begin{eqnarray}
\delta\psi_\mu&=&\left[\nabla_\mu-{1\over8}{\cal H}_{\mu\nu\lambda}
\gamma^{\nu\lambda}+{1\over2\sqrt{2}}(\overline{V}_R\overline{L}
{\cal F})_{\mu\nu}^{\overline a}\gamma^\nu\Gamma^{\overline{a}}
+\cdots\right]\epsilon\nonumber\\
\delta\lambda&=&-{1\over4\sqrt{2}}\left[\gamma^\mu\partial_\mu\eta
-{1\over6}{\cal H}_{\mu\nu\lambda}\gamma^{\mu\nu\lambda}
+{1\over2\sqrt{2}}(\overline{V}_R\overline{L}
{\cal F})_{\mu\nu}^{\overline{a}}\gamma^{\mu\nu}\Gamma^{\overline{a}}
+\cdots\right]\epsilon\ ,
\end{eqnarray}
Nester's form may be expressed as
\begin{eqnarray}
N^{\mu\nu}&=&\overline{\epsilon'}\gamma^{\mu\nu\rho}
\delta_{\epsilon}\hat\psi_\rho\nonumber\\
&=&\overline{\epsilon'}\gamma^{\mu\nu\rho}
\Bigl[\nabla_\rho+{1\over24}e^{-\eta}{\cal H}_{\eta\lambda\sigma}
(\gamma_\rho\gamma^{\eta\lambda\sigma}-3\delta_\rho{}^\eta
\gamma^{\lambda\sigma})\nonumber\\
&&\qquad\qquad\quad-{1\over8\sqrt{2}}e^{-\eta/2}
(\overline{V}_R\overline{L} {\cal F})_{\lambda\sigma}^{\overline{a}}
(\gamma_\rho\gamma^{\lambda\sigma} -4\delta_\rho{}^\lambda\gamma^\sigma)
\Gamma^{\overline{a}}+\cdots\Bigr]\epsilon\nonumber\\
&=&N_0{}^{\mu\nu}+{1\over2\sqrt{2}}e^{-\eta/2}\overline{\epsilon'}
(\overline{V}_R\overline{L}({\cal F}-i\gamma^5*{\cal F})^{\mu\nu}
)^{\overline{a}}\Gamma^{\overline{a}}\epsilon+\cdots\ .
\end{eqnarray}
In the last line, $N_0{}^{\mu\nu}$ is Nester's original expression
\cite{Nester}, which gives the ADM mass when integrated over the boundary
at spatial infinity
\begin{equation}
\overline{\epsilon'}P_\mu\gamma^\mu\epsilon={1\over4\pi G}
\int_{S^2_\infty}*N_0\ .
\end{equation}
 Defining the charges by the asymptotic behavior of the gauge fields
\begin{equation}
{\cal F}_{0r}\sim {Q\over r^2}\qquad *{\cal F}_{0r}\sim {P\over r^2}\ ,
\end{equation}
the surface integral of Nester's form gives
\begin{equation}
{1\over4\pi G}\int_{S^2\infty}*N=\overline{\epsilon'}\left[P_\mu\gamma^\mu
+{1\over2\sqrt{2}G}e^{-\eta_0/2}(\overline{V}_R\overline{L}
(Q-i\gamma^5P))^{\overline{a}}\Gamma^{\overline{a}}\right]\epsilon\ .
\end{equation}
Either application of the supersymmetry algebra or explicit calculation
then insures that this expression must be non-negative (provided the
equations of motion are satisfied).  From a four-dimensional $N=4$ point
of view, the Bogomol'nyi bound may then be written%
\begin{equation}
M\ge |Z_1|,|Z_2|\ ,
\end{equation}
where%
\footnote{These central charges have been noted independently by Cveti\v c
and Youm in \cite{Cvetic}.  Note, however, that our Nester procedure does
not yield the extra charge constraint found in \cite{Cvetic} on the basis
of black hole solutions.}
\begin{equation}
|Z_{1,2}|^2={1\over(4G)^2}e^{-\eta_0}\left[Q_R^2+P_R^2\pm
2\left(Q_R^2P_R^2-(Q_RP_R)^2\right)^\half\right]\ .
\end{equation}
The six right-handed electric charges are given by
\begin{equation}
Q_R^{\overline{a}}=\sqrt{2}(\overline{V}_R\overline{L}Q)^{\overline{a}}\ ,
\end{equation}
(and similarly for $P_R$).
This generalizes the Bogomol'nyi bound of \cite{Harveyliu}, which
holds only when the two central charges are identical, $|Z_1|=|Z_2|$.

Note that by using (\ref{eq:vlvr}), the square of
the right handed charges may be expressed as the $O(6,22;\BbbZ)$ invariant
combination
\begin{equation}
Q_R{}^2=Q^T\overline{L}(\overline{M}+\overline{L})\overline{L}Q\ .
\end{equation}
This allows us to write the central charges as
\be
|Z_{1,2}|^2=\frac{1}{16G^2}
\left[\g_{ia} {\cal M}_{Sij}(\overline{M}+\overline{L})_{ab}
\g_{jb}
 \pm  \sqrt{(\g_{ia}\e_{ij}\g_{jb})
    (\g_{kc}\e_{kl}\g_{ld})(\overline{M}+\overline{L})_{ac}
    (\overline{M}+\overline{L})_{bd}}
     \right]\ ,
\label{centralcharge}
\ee
where the electric and magnetic charges have been combined into a single
$SL(2,\BbbZ)\times O(6,22;\BbbZ)$ vector
\begin{equation}
\gamma_{ia}=\pmatrix{\a_S^a \cr \b_S^a}=\pmatrix{
         e^{-\eta_0}\overline{M}^{-1} & -a_{(0)} \overline{L} \cr
            0 & \overline{L}}^{ab} \pmatrix{Q\cr P}^b.
\end{equation}
The first feature to notice is that they are manifestly $SL(2,\BbbZ)_S$
invariant which is of relevance for $S$-duality invariance of heterotic
string theory. It is a well-known fact \cite{Sen7} that the spectrum of
states in the short $N=4$ multiplets is $SL(2,\BbbZ)_S$ invariant. In that case
$|Z_1|=|Z_2|$ and we recover from (\ref{centralcharge}) the Schwarz-Sen
formula
\be
M^2=\frac{1}{16G^2}
\g_{ia} {\cal M}_{Sij}(\overline{M}+\overline{L})_{ab}
\g_{jb}\ .
\label{schwsen}
\ee
However, a
discussion for the intermediate multiplets was missing so far.  The
masses of the states in those multiplets are given by $m={\rm Max}
(|Z_1|,|Z_2|)$.  Due to the familiar nonrenormalization theorems the
central charges do not receive any quantum corrections which also implies
that the masses are not renormalized.  $S$-invariance of
(\ref{centralcharge}) now gives the expected result that the full
supersymmetric mass spectrum has that property.

For the truncated set of fields considered in section (\ref{sec:N2Bog}), we
return to the notation of right-handed charges $Q_R$ and $P_R$.  If only
charges 1 and 2 are active, the central charges then reduce to
\begin{eqnarray}
(4G)^2|Z_1|^2&=&e^{-\eta_0}[(Q_R{}^1+P_R{}^2)^2+(Q_R{}^2-P_R{}^1)^2]
\nonumber\\
(4G)^2|Z_2|^2&=&e^{-\eta_0}[(Q_R{}^1-P_R{}^2)^2+(Q_R{}^2+P_R{}^1)^2]\ .
\label{eq:H2Bog}
\end{eqnarray}
This corresponds to the mass bound (\ref{us}) of section (\ref{sec:N2Bog}),
and agrees with the formula of \cite{Kallosh,Kalloshpeet}.

Now we are ready to repeat the analysis of section (\ref{sec:N2Bog}) for the
various black hole types. Again we choose vanishing background.
For dilaton couplings $a=\sqrt{3}$ and $a=1$
the square root term vanishes which implies $|Z_1|=|Z_2|$ and
(\ref{centralcharge})
reduces to the Schwarz-Sen mass formula. It was shown in \cite{Rahmfeld1}
that both black holes satisfy that Bogomol'nyi bound and therefore preserve
1/2 of the supersymmetries in $N=4$.
What happens to the other two black holes when embedded in the $N=4$ theory?
For the $a=1/\sqrt{3}$ black hole with charge vectors as given in
section (\ref{sec:N2Bog}) (the additional 24 electric and 24 magnetic charges
are zero) we find $|Z_1|=3/4$ and $|Z_2|=1/4$. With the
knowledge that the mass was given by $m=3/4$ we conclude that this
state preserves only one supersymmetry in $N=4$. This also holds
for dilaton coupling $a=0$. Here we find $m=|Z_1|=1$, $Z_2=0$, leading to the
same supersymmetry structure. Both black holes are in intermediate multiplets
of the $N=4$ supersymmetry algebra. All four values of $a$ yield special
cases of the general solutions recently found in \cite{Cvetic}.

\bigskip

It is also instructive to examine the Bogomol'nyi mass bound from the model
$A$ point of view.  In this case we start with the supersymmetry variation
of the four-dimensional Type $IIA$ gravitino
\begin{eqnarray}
\delta\skew5\hat{\tilde\psi}_\mu&=&\Bigl[\nabla_\mu+{1\over24}e^{-\tilde\eta}
\tilde {\cal H}_{\eta\lambda\sigma}\Gamma^{\hat7}
(\gamma_\mu\gamma^{\eta\lambda\sigma}-3\delta_\mu{}^\eta
\gamma^{\lambda\sigma})\\
&&\ +{1\over16}(e^{-\tilde\eta/2}(\tilde G_{ij}\tilde F_{\lambda\sigma}^j
+\tilde{\cal H}_{\lambda i\sigma}\Gamma^{\hat7})\Gamma^i
+\sqrt{2}e^{-\tilde\sigma/2}(\tilde V_R\tilde L
\tilde {\cal F})_{\lambda\sigma}^{\overline{a}}\Gamma^{\overline{a}})
(\gamma_\mu\gamma^{\lambda\sigma}-4\delta_\mu{}^\lambda\gamma^\sigma)
+\cdots\Bigr]\epsilon\ .\nonumber
\end{eqnarray}
This gives for Nester's expression
\begin{eqnarray}
\tilde N^{\mu\nu}&=&\tilde N_0{}^{\mu\nu}
+\overline{\epsilon'}\Biggl[{1\over4}e^{-\tilde\eta/2}
\Bigl((\tilde G_{ij}\tilde F^{j\mu\nu}+\epsilon_{ij}
*\tilde{\cal H}^{\mu j\nu})-i\gamma^5
(\tilde G_{ij}*\tilde F^{j\mu\nu}-\epsilon_{ij}
\tilde{\cal H}^{\mu j\nu})\Bigr)\Gamma^i
\nonumber\\
&&\kern10em+{1\over2\sqrt{2}}e^{-\tilde\sigma/2}
(\tilde V_R\tilde L(\tilde{\cal F}
-i\gamma^5*\tilde{\cal F})^{\mu\nu})^{\overline{a}}\Gamma^{\overline{a}}
\Bigr]\epsilon\ .
\label{eq:ANes}
\end{eqnarray}
This shows that, as far as the six-dimensional gauge fields are concerned,
the Type $IIA$ mass bound is identical to that of the Heterotic string.
Indeed, since the $S$--$T$ interchange is only applicable to the $6\to4$
fields, only their contributions to the Bogomol'nyi bound are modified.

{}From (\ref{eq:ANes}) we see that the four charges coming from the
compactification on $T^2$ enter into the mass formula in the combinations
\begin{eqnarray}
\tilde Q^a&=&\tilde Q_G^a+\epsilon^a{}_b\tilde P_B^b\nonumber\\
\tilde P^a&=&\tilde P_G^a-\epsilon^a{}_b\tilde Q_B^b\ ,
\end{eqnarray}
where $\tilde Q_G$ and $\tilde Q_B$ are defined by the asymptotic behavior
\begin{eqnarray}
\tilde E_i{}^a\tilde F^i_{0r}&\sim&{\tilde Q_G^a\over r^2}\nonumber\\
\tilde E^i{}_a\tilde{\cal H}_{0ir}&\sim&{\tilde Q_B^a\over r^2}
\end{eqnarray}
($E$ is the 4,5 components of the vierbein) and similarly for $\tilde P_G$
and $\tilde P_B$.  The two central charges are then given by
\begin{equation}
|\tilde Z_{1,2}|^2={1\over(4G)^2}\left[\tilde{\cal Q}^2+\tilde{\cal P}^2
\pm2\left(\tilde{\cal Q}^2\tilde{\cal P}^2-(\tilde{\cal Q}
\tilde{\cal P})^2\right)^{1\over2}\right]\ ,
\end{equation}
where we have grouped the 6 electric charges according to
\begin{equation}
\tilde{\cal Q}=[e^{-\tilde\eta/2}\tilde Q^a\quad
e^{-\tilde\sigma/2}\tilde Q_R^{\overline{a}}]^T\ .
\end{equation}
The right-handed charges $\tilde Q_R^{\overline{a}}$ are related to the
charges carried by the six-dimensional gauge fields
\begin{equation}
\tilde Q_R^{\overline{a}}
=\sqrt{2}(\tilde V_R\tilde L\tilde Q_F)^{\overline{a}}\ ,
\end{equation}
and correspond exactly to their heterotic counterparts
($\tilde Q_R^{\overline{a}}=Q_R^{\overline{a}}$ for
$\overline{a}=6,\ldots9$). Analogous definitions hold
for $\tilde{{\cal P}}$.

For vanishing $\tilde Q_R^{\overline{a}}$, the central charges become
\begin{eqnarray}
(4G)^2|\tilde Z_1|^2&=&e^{-\tilde\eta_0}[(\tilde Q_R{}^1+\tilde P_R{}^2)^2
+(\tilde Q_R{}^2-\tilde P_R{}^1)^2]\nonumber\\
(4G)^2|\tilde Z_2|^2&=&e^{-\tilde\eta_0}[(\tilde Q_L{}^1-\tilde P_L{}^2)^2
+(\tilde Q_L{}^2+\tilde P_L{}^1)^2]\ ,
\label{eq:A2Bog}
\end{eqnarray}
where the $6\to4$ charges are grouped into the combination
\begin{eqnarray}
\tilde Q_R{}^a&=&\tilde Q_G^a+\tilde Q_B^a\nonumber\\
\tilde Q_L{}^a&=&\tilde Q_G^a-\tilde Q_B^a\ .
\end{eqnarray}

For the Type $IIB$ string, we once again start with the four-dimensional
gravitino variation
\begin{equation}
\delta\skew6\hat{\skew5\tilde{\tilde{\psi}}}_\mu=\left[\nabla_\mu
-{1\over16}\Gtt_{ij}\Ftt_{\lambda\sigma}^j
(\gamma_\mu\gamma^{\lambda\sigma}-4\delta_\mu{}^\lambda\gamma^\sigma)\Gamma^i
-{1\over16}\cFtt_{i\,\lambda\sigma}^{a+}
(\gamma_\mu\gamma^{\lambda\sigma}+4\delta_\mu{}^\lambda\gamma^\sigma)
\Gamma^iT^a\right]P_R\epsilon\ .
\end{equation}
Since the spinors are chiral in six dimensions, we have explicitly
inserted the projection $P_R={1\over2}(1+\Gamma^{\hat7})
={1\over2}(1+\gamma^5\Gamma^{\hat3})$ into the above.  Taking into account
the self-duality of $\cFtt^+$, we arrive at
\begin{equation}
\Ntt^{\mu\nu}=\Ntt_0^{\mu\nu}
+\overline{\epsilon'}\left[
{1\over4}\Gtt_{ij}(\Ftt^{j\,\mu\nu}-i\gamma^5*\Ftt^{j\,\mu\nu})\Gamma^i
-{1\over4}(\cFtt_i^{a+\,\mu\nu}
-i\gamma^5*\cFtt_i^{a+\,\mu\nu})\Gamma^i
T^a\right]P_R\epsilon+\cdots\ .
\end{equation}
In this picture it is natural to define the Kaluza-Klein electric and
magnetic charges
\begin{equation}
\Ftt^i_{0r}\sim {\Qtt^i\over r^2}\qquad
*\Ftt^i_{0r}\sim {\Ptt^i\over r^2}\ .
\end{equation}
For the remaining gauge fields, we may define the $2\times26$ charges
\begin{equation}
\cFtt^{a+}_{i\,0r}\sim {\overline{Q}^a_i\over r^2}\ .
\end{equation}
Self-duality then gives the relation between ``electric'' and ``magnetic''
charges, $\overline{Q}_i^a= \epsilon_i{}^j\overline{P}_j^a$.  With these
definitions, the central charges in model $B$ have the form
\begin{eqnarray}
|\Ztt_{1,2}|^2&=&{1\over(4G)^2}\Biggl[
(\Qtt^i+\epsilon^i{}_j\Ptt^j)^2
+2(\overline{Q}^a\!\cdot\overline{Q}^a
+\overline{P}^a\!\cdot\overline{P}^a)\nonumber\\
&&\quad\pm2\left(4(\Ptt\cdot\overline{P}^a
+\Qtt\cdot\overline{Q}^a)^2
+2(\overline{Q}^a\!\cdot\overline{P}^b\overline{Q}^a\!\cdot\overline{P}^b
-\overline{Q}^a\!\cdot\overline{P}^b \overline{Q}^b\!\cdot\overline{P}^a)
\right)^{1\over2}\Biggr]\ .
\end{eqnarray}
The contractions denoted by $\cdot$ are over $i=4,5$ and are done with the
metric $\Gtt$.

For the truncated models of section (\ref{sec:N2D4}), only one of the
six-dimensional fields is active.  In this case, the two central charges
reduce to
\begin{equation}
|\Ztt_{1,2}|^2={1\over(4G)^2}\sum_{i=4,5}
\left[\Qtt_i +\epsilon_i{}^j\Ptt_j\pm
(\overline{Q}_i+\epsilon_i{}^j\overline{P}_j)\right]^2\ .
\label{eq:BbogZ}
\end{equation}
As previously, we denote left- and right-handed charges (with the vierbein
removed) in the combinations
\begin{eqnarray}
\Qtt_{R,L}&=&\Ett\Qtt
\pm\Ett^{-1}\overline{Q}\nonumber\\
\Ptt_{R,L}&=&\Ett\Ptt
\pm\Ett^{-1}\overline{P}\ ,
\end{eqnarray}
so that the central charges of (\ref{eq:BbogZ}) may be written
\begin{eqnarray}
(4G)^2|\Ztt_1|&=&(\Qtt_R{}^1+\Ptt_R{}^2)^2
+(\Qtt_R{}^2-\Ptt_R{}^1)^2\nonumber\\
(4G)^2|\Ztt_2|&=&(\Qtt_L{}^1+\Ptt_L{}^2)^2
+(\Qtt_L{}^2-\Ptt_L{}^1)^2\ .
\label{eq:B2Bog}
\end{eqnarray}
Compared to (\ref{eq:H2Bog}) the charges have no dilaton prefactor since they
have been defined canonically.  This completes the identification of the
central charges in all three models.

The central charges of the truncated theories, as given by (\ref{eq:H2Bog}),
(\ref{eq:A2Bog}) and (\ref{eq:B2Bog}), are summarized in Table 2.
Naturally, in the heterotic ($S$) language we verify the result of
\cite{Harveyliu} that only the right-handed charges contribute to the
central charges.  From the Type $II$ point of view we find a democracy
between right- and left-handers.  Each handedness goes along with one
central charge.
Naturally, the same result is obtained by dualizing the central charges
of the heterotic string. This implies that the dual of the $N=4$
heterotic string must be a Type $II$ string.

\begin{table}
$$
\begin{array}{cc}
\rm string&\rm central\ charge\\
&\\
S\hbox{--string}&Z_1{}^2=(Q_R{}^1+P_R{}^2)^2+(Q_R{}^2-P_R{}^1)^2\\
&Z_2{}^2=(Q_R{}^1-P_R{}^2)^2+(Q_R{}^2+P_R{}^1)^2\\
T\hbox{--string}&{\tilde Z}_1{}^2=(\tilde Q_R{}^1+\tilde P_R{}^2)^2
+(\tilde Q_R{}^2-\tilde P_R{}^1)^2\\
&{\tilde Z}_2{}^2=(\tilde Q_L{}^1-\tilde P_L{}^2)^2
+(\tilde Q_L{}^2+\tilde P_L{}^1)^2\\
U\hbox{--string}&{\Ztt}_1{}^2=(\Qtt_R{}^1
+\Ptt_R{}^2)^2
+(\Qtt_R{}^2-\Ptt_R{}^1)^2\\
&{\Ztt}_2{}^2=(\Qtt_L{}^1+\Ptt_L{}^2)^2
+(\Qtt_L{}^2-\Ptt_L{}^1)^2
\end{array}
$$
\caption{Central charges for the three theories.  We have removed a
prefactor of $4G$ as well as the asymptotic value of the dilaton field.}
\label{table3}
\end{table}

Although the physical states of all three strings must be identical as a
condition for string/string/string triality, the interpretation of the
spectrum in terms of elementary versus solitonic excitations is different in
the heterotic and Type $II$ theories (in $D=4$ the $IIA$ and $IIB$ {\it
elementary}
massive spectra
have identical interpretations).  In order to examine the elementary string
excitations, we set all magnetic charges to zero in the mass bound.
For the truncated heterotic theory, Table 2 gives
\begin{equation}
|Z_1|^2=|Z_2|^2={1\over(4G)^2}e^{-\eta_0}[(Q_R{}^1)^2+(Q_R{}^2)^2]\ ,
\end{equation}
which indicates that all Bogomol'nyi saturated elementary states in the
heterotic theory fall into short multiplets.  For the NS sector of the
heterotic string, the mass formula for string states,
$M^2=L_0=\overline{L}_0$, becomes
\begin{eqnarray}
M^2={1\over{16G^2}}e^{-\eta_0}[(Q_L)^2+(N_L-1)]
&=&{1\over16{G^2}}e^{-\eta_0}[(Q_R)^2+(N_R-\half)]\nonumber\\
&=&|Z_1|^2+{1\over{16G^2}}e^{-\eta_0}[(N_R-\half)]\ ,
\end{eqnarray}
giving the well-known result that the elementary heterotic states saturating
the Bogomol'nyi bound must satisfy $N_R={1\over2}$ \cite{Sen6,Rahmfeld1}.

On the other hand, from a Type $II$ point of view, the central charges are
given by
\begin{equation}
|\tilde Z_1|^2={1\over(4G)^2}e^{-\tilde\eta_0}
[(\tilde Q_R{}^1)^2+(\tilde Q_R{}^2)^2]\qquad
|\tilde Z_2|^2={1\over(4G)^2}e^{-\tilde\eta_0}
[(\tilde Q_L{}^1)^2+(\tilde Q_L{}^2)^2]\ .
\end{equation}
Thus the elementary Type $II$ string excitations saturating the Bogomol'nyi
bound may fall in either short or intermediate representations depending on
whether $(\tilde Q_L)^2=(\tilde Q_R)^2$ or not.  The Type $II$ string
mass formula in the NS-NS sector is%
\footnote{Space-time bosons in the R-R sector satisfy a similar equation.
While no elementary string states carry R-R charge, states from the R-R
sector may be charged under the NS-NS gauge bosons.}
\begin{eqnarray}
M^2&=&{1\over(4G)^2}e^{-\tilde\eta_0}[(\tilde Q_L)^2+(\tilde N_L-\half)]
={1\over(4G)^2}e^{-\tilde\eta_0}[(\tilde Q_R)^2+(\tilde N_R-\half)]
\nonumber\\
&=&|\tilde Z_2|^2+{1\over(4G)^2}e^{-\tilde\eta_0}
[(\tilde N_L-\half)]=|\tilde Z_1|^2+{1\over(4G)^2}
            e^{-\tilde\eta_0}[(\tilde N_R-\half)]\ .
\end{eqnarray}
This indicates that Bogomol'nyi states are in short multiplets for
$\tilde N_L=\tilde N_R={1\over2}$ and intermediate multiplets for $\tilde
N_L>\tilde N_R={1\over2}$ or $\tilde N_R>\tilde N_L={1\over2}$.

\section{String and fivebrane solitons}
\label{sec:N4Sol}

When the full set of fields are included, one may once again find the
three string soliton solutions of section (3) but now the zero-mode
structures will be more complicated. Ideally, in fact, one would like
them to correspond to the worldsheet field content of the heterotic,
Type $IIA$ and Type $IIB$ superstrings.

That the Type $IIA$ theory in $D=6$ admits a soliton with the correct
heterotic zero-modes was discussed in \cite{Senssd,Harvey}. Just as we
found the 4-parameter deformation in section (5)
by making $O(2,1)/O(2) \times
O(2,1)/O(2)$ transformations on the neutral solution so we may find the
extra 24 parameters by making $O(20,1)/O(20) \times O(4,1)/O(4)$
transformations. When combined with the translation modes and their
fermionic partners, one finds in this way for the physical degrees of
freedom a total of $8$ right moving bosons, $8$ right moving fermions
and $24$ left moving bosons appropriate to the fundamental heterotic
string \cite{Senssd}.  In fact, the same result may be obtained
\cite{Harvey,Townsendseven,Duffliu} by starting with the physical
zero modes of the Type $IIA$ fivebrane soliton in $D=10$ \cite{Callan2},
namely the $d=6$ chiral supermultiplet
$(B^-{}_{\mu\nu},\lambda^I,\phi^{[IJ]})$, and wrapping the fivebrane
around $K3$ \cite{Minasian}.

Finding the Type $II$ strings as solitons of the heterotic string is
more problematical, however. Although the zero modes associated with
the $4$ NS charges may be obtained in the same way, this is not true of
the $24$ RR charges since the fundamental Type $II$ strings do not
carry these charges \cite{Senssd,Harvey}.  The problem of identifying
these zero modes is akin to the missing monopole problem
\cite{Gauntlett} and requires a better understanding of the
role of $K3$ in counting the dimension of the moduli space.

Since the Type $IIA$/heterotic duality admits a $D=10$ fivebrane
interpretation, one might expect the same to be true of Type $IIB$ now
that it has been included in the picture via four dimensional
string/string/string triality.  However, in this case the critical
solitonic string found in $D=4$ does not seem to be related to the
$D=6$ string obtained by wrapping the $D=10$ fivebrane around $K3$
since this latter string appears not to be critical \cite{Townsendseven}.
This is in need of further study.

\section{Conclusion}
\label{sec:Conclusion}

{}From one point of view, four-dimensional string/string/string triality
seems a trivial extension of what we already knew: $D=6$ string/string
duality accompanied by mirror symmetry. Yet, as we have seen, it has
far-reaching consequences. $D=6$ string/string duality satisfactory accounts
for strong/weak coupling duality of the Type $IIA$ string in terms of
$SL(2,\BbbZ)_T$,
the target space duality of the heterotic string, but leaves a gap in
accounting for the converse, because $SL(2,\BbbZ)_S$ takes R-R fields
of Type $IIA$ into their duals. Four-dimensional string/string/string duality
fills this gap: $SL(2,\BbbZ)_S$ is guaranteed by $D=6$ general covariance
of the Type $IIB$ string. Moreover, since the conjectured $SL(2,\BbbZ)_X$
of the Type $IIB$ string is just a subgroup of the $O(6,22;\BbbZ)_{TU}$
target space duality of the heterotic string, we see that this triality also
accounts for this symmetry and hence for {\it all} the conjectured
non-perturbative symmetries of string theory.

\bigskip

\bigskip
\noindent
{\Large {\bf Acknowledgements}}

It is a pleasure to thank Ashoke Sen for useful conversations.

\bigskip

\bigskip
\noindent
{\Large {\bf Note Added}}

After the completion of this work, we became aware of a paper by
Girardello, Porrati and Zaffaroni \cite{PorrHet}, which also
displays the $D=4$ heterotic/$IIA$ dictionary and also discusses
the absence of a perturbative $T$-duality in the Type $IIA$ theory and
hence a gap in deriving $S$-duality of the heterotic string from
$D=6$ string/string duality \cite{Duffstrong} alone. However,
this gap is filled by the $D=4$
string/string/string triality of the present paper: $SL(2,\BbbZ)_S$
is guaranteed by $D=6$ general covariance of the Type $IIB$ string.

\newpage

\appendix
\section{Appendix}

In this appendix we examine the compactifications of ten-dimensional string
theories that give rise to the six-dimensional models of section
(\ref{sec:N2D6}).  For the first case, we consider the heterotic string
compactified on $T^4$, giving rise to model $H$.  A toroidal
compactification is straightforward, and gives rise to the action
(\ref{eq:Hfull}).  As far as the bosonic fields are concerned, all that
remains is to specify the $O(4,20)$ matrix $M$.  This matrix may be
decomposed in terms of a vierbein, $M=V^TV$ where $V$ transforms as a
vector under both $O(4,20;\BbbZ)$ and $O(4)\times O(20)$ and satisfies
\begin{equation}
V^{-1}=[\eta VL]^T\ ,
\end{equation}
where
\begin{equation}
\eta=\pmatrix{I_4&0\cr 0&-I_{20}}\ .
\end{equation}
In terms of the original ten dimensional heterotic fields, the vierbein may
be written as
\begin{equation}
V^{\overline{a}}{}_b =\left[\matrix{V_R\cr V_L}\right]
=\left[\matrix{{1\over\sqrt{2}}E^{-1}\cr&{1\over\sqrt{2}}E^{-1}\cr
&&I_{16}}\right]\times
\left[\matrix{I_4&(G+B+C)&-A\cr I_4&(-G+B+C)&-A\cr 0&A^T&-I_{16}}\right]\ ,
\end{equation}
where the 24 gauge fields have been arranged in the order of 4
Kaluza-Klein, 4 winding, and 16 heterotic $U(1)$'s (see {\it e.g.}\
Ref.~\cite{Sen2,Sen6}). $V_R$ and $V_L$ denotes the split of the vierbein
into right- and left-handed components transforming under $O(4)$ and
$O(20)$ respectively and satisfies
\begin{equation}
V_L^TV_L^{\vphantom{T}}=\half(M-L)\qquad V_R^TV_R^{\vphantom{T}}=\half(M+L)\ .
\label{eq:vlvr}
\end{equation}
%


We now turn to the compactification of $D=10$ Type $II$ strings to six
dimensions.  Since the compactifications of interest involve $K3$, we first
list some of its important properties.  The Betti numbers are given by
$b_0=1$, $b_1=0$, $b_2^+=3$ and $b_2^-=19$, so
we may choose an integral basis of harmonic two-forms, $\omega_2$, with
intersection matrix
\begin{equation}
d_{IJ}=\int_{K3}\omega_I\wedge\omega_J\ .
\end{equation}
Since taking a Hodge dual of $\omega_I$ on $K3$ gives another harmonic
two-form, we may expand the dual in terms of the original basis
\begin{equation}
*\omega_I=\omega_JH^J{}_I\ .
\end{equation}
In this case, we find
\begin{equation}
\int_{K3}\omega_I\wedge*\omega_J=d_{IK}H^K{}_J\ .
\end{equation}
The matrix $H^I{}_J$ depends on the metric on $K3$, and hence the
$b_2^+\cdot b_2^-=57$ $K3$ moduli.  Because $**=1$, $H^I{}_J$ satisfies
the properties
\begin{eqnarray}
H^I{}_JH^J{}_K&=&\delta^I{}_K\nonumber\\
d_{IJ}H^J{}_K&=&d_{KJ}H^J{}_I\ ,
\end{eqnarray}
so that
\begin{equation}
H^J{}_Id_{JK}H^K{}_L=d_{IL}\ .
\end{equation}
Since $H^I{}_J$ has eigenvalues $\pm1$, it may be diagonalized by a
similarity transformation
\begin{equation}
O^i{}_JH^J{}_K(O^{-1})^K{}_l=\eta^i{}_l
\qquad H^I{}_J=(O^{-1})^I{}_k\eta^k{}_lO^l{}_J\ ,
\end{equation}
where $\eta$ has signature $(3,19)$.
Using $O(3)\times O(19)$ invariance, we may always choose $O$ such that
\begin{eqnarray}
d_{IJ}&=&O^k{}_I\eta_{kl}O^l{}_J\nonumber\\
d^{IJ}&=&(O^{-1})^I{}_k\eta^{kl}(O^{-1})^J{}_l\ ,
\end{eqnarray}
where $d^{IJ}$ is the inverse of $d_{IJ}$.

For the Type $IIA$ supergravity compactified on $K3$, the ten-dimensional
3-form potential gives rise to 22 six-dimensional gauge fields and a
remaining 3-form which may be dualized as mentioned in the previous
discussion.  These 23 gauge fields, plus another originating from the
1-form potential in ten dimensions, enter into (\ref{eq:Afull}) with
$\tilde M$ given by a vierbein, $\tilde M=\tilde V^T\tilde V$ where
\begin{equation}
\tilde V^i{}_J=\left[\matrix{{1\over\sqrt{2}}e^{\rho/2}\cr&{1\over\sqrt{2}}
e^{\rho/2}\cr&&I_{22}}\right]\times\left[\matrix{
-1&e^{-\rho}+\half(b^Ib^Jd_{IJ})&b^J\cr
1&e^{-\rho}-\half(b^Ib^Jd_{IJ})&-b^J\cr
0&O^i{}_Ib^I&O^i{}_Kd^{KJ}}\right]\ .
\end{equation}
The $O^i{}_J$ contain the 57 $K3$ moduli, $e^\rho$ is the
breathing mode, and the 22 $b^I$ correspond to torsion on $K3$.  This vierbein
satisfies
\begin{equation}
\tilde V^{-1}=[\tilde\eta\tilde V\tilde L]^T\ ,
\end{equation}
where
\begin{equation}
\tilde L=\pmatrix{\sigma^1&0\cr0&d_{IJ}}\ ,
\end{equation}
and
\begin{equation}
\tilde\eta=\pmatrix{-1\cr&1\cr&&\eta_{ij}}\ .
\end{equation}

In ten dimensions, the Type $IIB$ string contains both a complex scalar and
a complex 3-form field-strength which transform under
$SL(2,\BbbZ)_X$.  While the complete theory contains a 4-form potential,
$D_4{}^+$, with self-dual field strength and hence does not admit a
conventional Lagrangian formulation, it is possible to write down a
truncated action where $D_4{}^+$ is absent.  In natural string coordinates,
the partial bosonic action is \cite{Bergshoeffduality}
\begin{eqnarray}
I_{D=10}&=&{1\over2\kappa_{10}{}^2}\int d^{10}x\sqrt{-G^{(10)}}
e^{-\Phi^{(10)}}\Bigl[R_{G^{(10)}}+(\partial_M\Phi^{(10)})^2
-{1\over12}(H^{(1)}_{MNP})^2\nonumber\\
&&\qquad\qquad +e^{\Phi^{(10)}}\Bigl(-{1\over2}(\partial_M\ell)^2
-{1\over12}(H^{(2)}_{MNP}-\ell H^{(1)}_{MNP})^2\Bigr)\Bigr]\ ,
\label{eq:tenIIB}
\end{eqnarray}
where $H_3^{(i)}=dB_2^{(i)}$.  From a supergravity point of view, $H^{(1)}$
and $H^{(2)}$ are indistinguishable due to the $SL(2,\BbbZ)_X$ symmetry.
In fact, the truncated action may be written more symmetrically in
canonical coordinates where a real dilaton need not be singled out.
However string theory indicates that there is a single dilaton as well as a
single real 3-form coming from the NS-NS sector of the string.  These
fields are labeled by $\Phi^{(10)}$ and $H^{(1)}$ in (\ref{eq:tenIIB}),
whereas $\ell$ and $H^{(2)}$ arise from the R-R sector.

In the absence of a covariant action, the full ten-dimensional equations
of motion for the bosonic fields are given by \cite{Bergshoeffduality}
\begin{eqnarray}
R_{MN}-\half G_{MN}^{(10)}R &=&\kappa_{10}{}^2T_{MN}\nonumber\\
\nabla^2\Phi^{(10)}&=&-{1\over2}R_{G^{(10)}}
+{1\over2}(\partial\Phi^{(10)})^2 +{1\over24}(H^{(1)})^2\nonumber\\
\nabla^2\ell&=&-{1\over6}H^{(1)}(H^{(2)}-\ell H^{(1)})\nonumber\\
d*((\ell^2+e^{-\Phi^{(10)}})H_3^{(1)}-\ell H_3^{(2)})&=&F_5H_3^{(2)}\nonumber\\
d*(H_3^{(2)}-\ell H_3^{(1)})&=&-F_5H_3^{(1)}\nonumber\\
F_5&=&*F_5\nonumber\\
dF_5&=&H_3^{(1)}H_3^{(2)}\ ,
\end{eqnarray}
where the stress tensor is
\begin{eqnarray}
T_{MN}&=&{1\over2\kappa_{10}{}^2}\Bigl[
-2(\partial_M\Phi^{(10)}\partial_N\Phi^{(10)}
-{1\over2}G^{(10)}_{MN}(\partial\Phi)^2)
+{1\over2}(H^{(1)}_{MPQ}H^{(1)}_N{}^{PQ}
-{1\over6}G^{(10)}_{MN}(H^{(1)})^2)\nonumber\\
&&\qquad +e^{\Phi^{(10)}}\Bigl((\partial_M\ell\partial_N\ell
-{1\over2}G^{(10)}_{MN}(\partial\ell)^2)
+{1\over2}((H^{(2)}-\ell H^{(1)})_{MPQ}
(H^{(2)}-\ell H^{(1)})_N{}^{PQ}\nonumber\\
&&\qquad\quad-{1\over6}G^{(10)}_{MN}(H^{(2)}-\ell H^{(1)})^2)
+{1\over2\cdot4!}(F_{MPQRS}F_N{}^{PQRS}
-{1\over2}G^{(10)}_{MN}F^2)\Bigr)\Bigr]\ .
\end{eqnarray}
$F_5=dD_4{}^++\half\epsilon^{ij}B_2^{(i)}dB_2^{(j)}$ is the self-dual field
strength of the Type $IIB$ theory.

We compactify this theory by decomposing the 2-form and 4-form
potentials in a basis of harmonic forms on $K3$
\begin{eqnarray}
B_2^{(i)}&=&B_2^{(i)}+\a'b^{(i)I}\omega_I\nonumber\\
D_4{}^+&=&D_4+\a' D_2^I\omega_I+\a'^2
d\ \omega_4\ .
\end{eqnarray}
Note that the self-duality condition for $D_4{}^+$ allows us to
eliminate $D_4$ in favor of $d$.  This also ensures that, of the
22 $D_2^I$, three are self-dual and 19 are anti-self-dual in $D=6$
\begin{equation}
F_3^I=*F_3^JH^I{}_J\ ,
\end{equation}
where $F_3^I=dD_2^I$.  Further decomposing
$H_3^{(i)}$ into chiral parts gives a total of 5 self-dual and 21
anti-self-dual 3-form field strengths
in six dimensions.  Hence the compactified theory
has the field content of a chiral supergravity multiplet
$(E_M{}^A,\psi_M^I,B_{MN}^{+IJ})$ coupled to 21 tensor multiplets
$(B_{MN}^-,\lambda^I,\phi^{IJ})$.

The part of the six-dimensional action containing $H_3^{(i)}$ may be
written covariantly
\begin{equation}
I_B^{H^{(i)}}={1\over4\kappa^2}\int\left[e^{-\Phitt}
H_3^{(1)}*H_3^{(1)}
+e^{-\rho}H_3^{(2')}*H_3^{(2')}\right]\ ,
\label{eq:h1h2}
\end{equation}
however the full theory has no covariant action.
In the above, $\Phitt$ is the six-dimensional dilaton,
$\Phitt=\Phi^{(10)}+\rho$ where $\rho$ fixes the size of $K3$
\begin{equation}
e^{-\rho}={1\over V}\int_{K3}*1\ .
\end{equation}
We have also defined the shifted $H_3^{(2')}$ field by
$H_3^{(2')}=H_3^{(2)}-\ell H_3^{(1)}$.

In order to incorporate all 26 chiral 3-forms, we examine
the Bianchi identities and equations of motion to identify the
``field strengths'' ${\cal H}_3$ satisfying $d{\cal H}_3=0$:
\begin{equation}
{\cal H}_3= [H^1\quad H^2\quad-H^3\quad H^4\quad H^I]^T\ ,
\end{equation}
where
\begin{eqnarray}
H^1&=&H_3^{(1)}\nonumber\\
H^2&=&e^{-\Phitt}*H_3^{(1)}-\ell e^{-\rho}*H_3^{(2')}
-(d+\alpha b^1b^2) (H_3^{(2')}+\ell H_3^{(1)})+b^{(2)I}F_3^Jd_{IJ}
+{1\over2}b^2b^2H_3^{(1)}\nonumber\\
H^3&=&H_3^{(2')}+\ell H_3^{(1)}\nonumber\\
H^4&=&e^{-\rho}*H_3^{(2')}+(a+(\alpha-1)b^1b^2)H_3^{(1)}
-b^{(1)I}F_3^Jd_{IJ}+{1\over2}b^1b^1(H_3^{(2')}+\ell H_3^{(1)})\nonumber\\
H^I&=&F_3^I+b^{(2)I}H_3^{(1)}-b^{(1)I}(H_3^{(2')}+\ell H_3^{(1)})\ .
\label{eq:h3fs}
\end{eqnarray}
We have used a short-hand notation where $b^ib^j=b^{(i)I}b^{(j)J}d_{IJ}$ and
$\alpha$ is an arbitrary parameter.

On the other hand, the natural (anti-)self-dual field strengths are
\begin{equation}
\Htt_3^\pm=[H_3^{1+}\quad H_3^{1-}\quad H_3^{2+}\quad H_3^{2-}
\quad F_3^{i\pm}]^T\ ,
\end{equation}
where
\begin{eqnarray}
H_3^{1\pm}&=&{1\over\sqrt{2}}e^{-\Phitt/2}
(H_3^{(1)}\pm*H_3^{(1)})\nonumber\\
H_3^{2\pm}&=&{1\over\sqrt{2}}e^{-\rho/2}(H_3^{(2)}\pm*H_3^{(2)})\nonumber\\
F_3^{i\pm}&=&O^i{}_JF_3^J\ .
\label{eq:h3pm}
\end{eqnarray}

These 3-forms are related by a vierbein
\begin{equation}
{\cal H}_3=(\Ltt^{-1})(\Vtt^{-1})
\Htt_3^\pm \qquad
\Htt_3^\pm=\Vtt\Ltt{\cal H}_3\ ,
\end{equation}
which depends on the 57+22+1 $K3$ moduli, $O^i{}_J, b^{(1)I}, e^{-\rho}$, and
the 22+3 additional scalars $b^{(2)I}, e^{-\Phitt}, \ell$, and
$d$.  The $O(5,21)$ matrix $\Ltt$ has been defined in
(\ref{eq:tildetildeL}).
Using (\ref{eq:h3fs}) and (\ref{eq:h3pm}), we find for the vierbein
\begin{eqnarray}
\Vtt^{iJ}&=&\left[\matrix{{1\over\sqrt{2}}e^{\Phitt/2}\cr
&{1\over\sqrt{2}}e^{\Phitt/2}\cr
&&{1\over\sqrt{2}}e^{\rho/2}\cr
&&&{1\over\sqrt{2}}e^{\rho/2}\cr
&&&&I_{16}}\right]\\
\times\ &&\kern-32pt\left[\matrix{
-1
&\!-(e^{-\Phitt}-a\ell-\alpha\ell b^1b^2+{1\over2}b^2b^2)
&\ell
&\!-(a+(\alpha-1)b^1b^2+{1\over2}\ell b^1b^1)
&\!-(b^{(2)J}-\ell b^{(1)J})
\cr
1
&\!-(e^{-\Phitt}+a\ell+\alpha\ell b^1b^2-{1\over2}b^2b^2)
&-\ell
&\!(a+(\alpha-1)b^1b^2+{1\over2}\ell b^1b^1)
&\!(b^{(2)J}-\ell b^{(1)J})
\cr
0
&(\ell e^{-\rho}+a+\alpha b^1b^2)
&1
&-(e^{-\rho}+{1\over2}b^1b^1)
&b^{(1)J}
\cr
0
&(\ell e^{-\rho}-a-\alpha b^1b^2)
&-1
&-(e^{-\rho}-{1\over2}b^1b^1)
&-b^{(1)J}
\cr
0
&O^i{}_Ib^{(2)I}
&0
&-O^i{}_Ib^{(1)I}
&O^i{}_Id^{IJ}
}\right]
\hskip-8pt\nonumber
\end{eqnarray}
with inverse given by
\begin{equation}
\Vtt^{-1}=[\etatt\Vtt\Ltt]^T\ .
\end{equation}

Finally, the $O(5,21)/O(5)\times O(21)$ matrix of scalars is given by
$\Mtt=\Vtt^T\Vtt$ and the 3-form equations of motion are given by
\begin{equation}
d\Htt_3^\pm=d\Vtt\Vtt^{-1}\Htt^\pm\ .
\end{equation}

\newpage

\bibliographystyle{preprint}



\end{document}